\documentclass[journal,twoside]{IEEEtran}
\usepackage{amsmath,amsfonts}
\usepackage{algorithmic}
\usepackage{algorithm}
\usepackage{stfloats}
\usepackage{url,balance}
\usepackage{cite}
\usepackage{microtype}
\usepackage{graphicx}
\usepackage{booktabs} 
\usepackage[normalem]{ulem}
\usepackage{hyperref}
\usepackage{amssymb}
\usepackage{mathtools}
\usepackage{amsthm}
\usepackage{acronym}
\usepackage{arydshln}
\usepackage{multirow}
\usepackage{pifont}
\usepackage{soul} 
\usepackage{xcolor}
\acrodef{DPS}{diffusion posterior sampling}
\acrodef{DAPS}{decoupled annealing posterior sampling}
\acrodef{NMF}{non-negative matrix factorization}
\acrodef{EM}{expectation-maximisation}
\acrodef{SNR}{signal-to-noise ratio}
\acrodef{ODE}{ordinary differential equation}
\acrodef{SDE}{stochastic differential equation}
\acrodef{STFT}{short-time Fourier Transform}
\acrodef{SNR}{signal-to-noise ratio}
\acrodef{SIR}{signal-to-interference ratio}
\acrodef{ESTOI}{extended short-term objective intelligibility}
\acrodef{PESQ}{perceptual evaluation of
speech quality}
\acrodef{SI-SDR}{scale-invariant signal-to-distortion ratio}
\acrodef{NISQA}{Non-Intrusive Speech Quality Assessment}
\acrodef{WER}{word error rate}
\acrodef{ASR}{Automatic Speech Recognition}
\acrodef{TDT}{Token-and Duration Transducer}
\acrodef{NFE}{number of function evaluations}

\usepackage{amsmath,amsfonts,bm}

\def\1{\bm{1}}

\def\rvc{{\mathbf{c}}}

\def\rvh{{\mathbf{h}}}

\def\rvn{{\mathbf{n}}}

\def\rvs{{\mathbf{s}}}

\def\rvv{{\mathbf{v}}}
\def\rvw{{\mathbf{w}}}
\def\rvx{{\mathbf{x}}}
\def\rvy{{\mathbf{y}}}
\def\rvz{{\mathbf{z}}}

\def\rmI{{\mathbf{I}}}

\DeclareMathAlphabet{\mathsfit}{\encodingdefault}{\sfdefault}{m}{sl}
\SetMathAlphabet{\mathsfit}{bold}{\encodingdefault}{\sfdefault}{bx}{n}

\def\gA{{\mathcal{A}}}

\def\gL{{\mathcal{L}}}

\def\gN{{\mathcal{N}}}

\def\sR{{\mathbb{R}}}

\newcommand{\E}{\mathbb{E}}

\newcommand{\xmark}{\ding{55}}%

\usepackage{umoline}
\usepackage[capitalize,noabbrev]{cleveref}
\begin{document}

\title{SSNAPS: Audio-Visual Separation of Speech and Background Noise with Diffusion Inverse Sampling\thanks{This work was supported in part by the “AUDIENCE:
Audio-Visual Analysis and Separation” Project, Data Science Program, Council
of Higher Education, Israel.}}

\author{Yochai Yemini,~\IEEEmembership{Student Member,~IEEE,}, Yoav Ellinson,
\\ Rami Ben-Ari, Sharon Gannot~\IEEEmembership{Fellow,~IEEE,} and Ethan Fetaya
\thanks{\sloppy Y. Yemini, Y. Ellinson, S. Gannot and E. Fetaya are with Bar-Ilan University, Israel,  e-mail: \texttt{\{yochai.yemini,yoav.ellinson,sharon.gannot,\\ethan.fetaya\}@biu.ac.il}; R. Ben-Ari is with OriginAI, Israel, email: \texttt{ramib-at-originai.co}.
}
}

\markboth{IEEE Transactions on Audio, Speech and Language Processing, June 2026}%
{Yemini \MakeLowercase{\textit{et al.}}: SSNAPS: Audio-Visual Separation of Speech and Background Noise
with Diffusion Inverse Sampling}


\maketitle

\begin{abstract}
This paper addresses the challenge of audio-visual single-microphone speech separation and enhancement in the presence of real-world environmental noise. Our approach is based on generative inverse sampling, where we model clean speech and ambient noise with dedicated diffusion priors and jointly leverage them to recover all underlying sources. To achieve this, we reformulate a recent inverse sampler to match our setting. 
We evaluate on mixtures of 1, 2, and 3 speakers with noise. {Our experiments show that despite being entirely unsupervised, our method consistently outperforms leading supervised baselines in \ac{WER} in most conditions, and in all multi-speaker ones}. We further extend our framework to handle off-screen speaker separation.  
Moreover, the high fidelity of the separated noise component makes it suitable for downstream detection of the acoustic scene. 
Project page and code: \url{https://github.com/yochaiye/ssnaps/}
\end{abstract}

\begin{IEEEkeywords}
speech separation, speech enhancement, diffusion model, inverse problems.
\end{IEEEkeywords}

\section{Introduction}
A central problem in speech processing is extracting clean speech from recordings made in noisy conditions. Over the past decade, this field has shifted towards neural network-based approaches. Most of these methods tackle noise suppression in a supervised manner: a neural network is trained to reconstruct clean speech from its noisy counterpart. Such supervised approaches can achieve outstanding performance on the specific tasks and conditions for which they were trained \cite{sepformer,tfgrid,flowavse,sgmse,storm,gao2021VisualVoice}. However, one of their key limitations is a lack of flexibility. When the acoustic environment changes, supervised models typically need to be retrained to handle the new conditions and prevent a drop in performance.

{In contrast, unsupervised speech processing \cite{wisdom2020mixit,bando2018vaenmf,sadeghi2020avvae,leglaive2020rvae} adopts a different strategy.} In particular, unsupervised generative frameworks train a prior speech model that captures the distribution of \textit{clean} speech signals. This prior is then used to infer the desired speech signal from a degraded observation. The key advantage of this approach is that the speech prior is decoupled from the observation model. Consequently, unsupervised speech estimation is motivated by its flexibility: it does not require training a dedicated model for each degradation type, can accommodate arbitrary speech and noise models, and can, in theory, handle an arbitrary number of speakers. This versatility enables a broad range of speech processing tasks to be addressed within a unified paradigm.

The main interest of this paper is unsupervised speech separation in the presence of background noise. In particular, we focus on algorithms that use diffusion models \cite{ho2020denoising,song2021scorebased,karras2022elucidating} as data priors. More specifically, our goal is to deploy the inverse problem framework, in which an estimate of the clean data is obtained by sampling from the posterior distribution conditioned on the observation. While this technique has been successfully applied to various speech processing tasks, including \cite{buddy,moliner2023solving,moliner2024blind,arraydps}, the literature on diffusion-based speech separation in the presence of background noise remains scarce \cite{nmf1,nmf2,davssnm,xavi}. This is particularly surprising, given the abundance of diffusion inverse samplers, e.g. \cite{dps,pi-gdm,daps,ip1,ip2,ip3,ip4}.

In this work, we present SSNAPS, standing for Separation of Speech and Noise with Annealed Posterior Sampling. We adopt the approach presented in \cite{davssnm} and use two diffusion priors in conjunction: a clean speech prior and an environmental noise prior. The two data priors are jointly utilised to discern speech from noise. The former is further harnessed to separate competing speech sources. We primarily focus on the audio-visual setting, i.e., we assume that for each speech-and-noise mixture, lip-region videos of all speakers are available. Unlike \cite{davssnm}, which used the \ac{DPS} \cite{dps} framework as its inverse sampler, we rely on \ac{DAPS} \cite{daps}, a recently proposed paradigm. As noted in \cite{daps}, despite their popularity, \cite{dps} and other methods that rely on directly solving a reverse diffusion \ac{SDE} or \ac{ODE} with the posterior score function suffer from a major shortcoming. Because they often use accurate \ac{SDE}/\ac{ODE} solvers that take very small steps, they unintentionally inhibit global adjustments to the signal during sampling, potentially resulting in inaccurate separation. In contrast, \ac{DAPS} circumvents this setback by deploying a mechanism that decouples two successive sampling steps and does not directly solve the reverse differential equation.

Our contribution in this paper is fourfold. Firstly, we reformulate \ac{DAPS} to address our scenario, enabling it to estimate multiple statistically independent signals from two distinct distributions. Secondly, since our approach is entirely unsupervised, the same speech prior can formally be used for speaker separation regardless of the number of speakers. While in \cite{davssnm} evaluations were conducted only on two-speaker mixtures with noise, we evaluate on noisy mixtures of 1, 2, and 3 speakers. Our experimental study shows that SSNAPS achieves the best \ac{WER} scores, even compared to top-tier \emph{supervised} counterparts. Thirdly, we propose an extension of SSNAPS that allows it to perform off-screen speech separation. To the best of our knowledge, we are the first to propose an inverse problem technique for this scenario. Lastly, we demonstrate that the high-fidelity separation of the ambient noise further enables scene classification.

\section{Background}
\label{sec:background}
\subsection{Score-Based Diffusion Models}
Diffusion models generate samples from a data distribution $p(\rvx)$ by reversing a predefined noising process, also called the forward process. {The forward pass takes a clean speech signal $\rvx\sim p(\rvx)$, denoted $\rvx_0$, and gradually adds increasing levels of Gaussian noise to it, such that at time step $\tau$ the noisy diffusion state $\rvx_\tau$ satisfies $p_\tau(\rvx_\tau\mid\rvx_0) = \gN(\rvx_0, \sigma_{\tau}^2\rmI)$.} The values of $\sigma_{\tau}$ are chosen such that $\sigma_{\tau}$ is monotonically increasing, where $\sigma_0=0$ and $\rvx_{T_{\max}}\sim \gN(\mathbf{0}, \sigma_{T_{\max}}^2\rmI)$. In \cite{karras2022elucidating}, $\sigma_{\tau}=\tau$ is chosen, and we follow this choice in our derivations. This incremental noise injection can be described \cite{karras2022elucidating} with the following \ac{SDE}:
\begin{equation}
    \mathrm{d}\rvx_\tau = \sqrt{2\dot{\sigma}_\tau\sigma_\tau}\mathrm{d}\rvw_\tau
\label{eq:sde}
\end{equation}
where, the smallest diffusion step is $\tau=T_{\min}$ and not $\tau=0$, for some arbitrarily small $T_{\min}$, to ensure numerical stability. The dot operation denotes a derivative with respect to $\tau$, and $\rvw_\tau$ is a standard Wiener process. 

Reversing (\ref{eq:sde}) and thus obtaining a sample $\rvx_0$ from $p(\rvx)$ can be accomplished by solving the following variance exploding \ac{ODE} \cite{karras2022elucidating}:
\begin{equation}
    \mathrm{d}\rvx_\tau = -\dot{\sigma}_{\tau}\sigma_{\tau}\nabla_{\rvx_\tau}\log p_\tau(\rvx_\tau)\mathrm{d}\tau.
\label{eq:ode}
\end{equation}
The \ac{ODE} transports a noise sample from $\gN(\mathbf{0}, \sigma_{T_{\max}}^2\rmI)$ to a sample $\rvx_0$ from $p(\rvx)$. The score function $\nabla_{\rvx_\tau}\log p(\rvx_\tau)$ is not directly computable at inference time because $\rvx_0$ is unavailable. Instead, it is approximated using a diffusion denoiser $D_{\theta}(\rvx_\tau, \tau)$, which is trained to predict $\hat{\rvx}_0 \approx \rvx_0$ by minimizing the following $L_2$ objective:
\begin{equation}
    \mathbb{E}_{\tau, \rvx\sim p(\rvx), \boldsymbol{\epsilon}\sim\gN(\mathbf{0},\rmI)}\left\|D_{\theta}(\rvx+\sigma_\tau\boldsymbol{\epsilon},\tau) - \rvx\right\|^2.
\end{equation}
Then, by invoking Tweedie's formula \cite{tweedies}, the score function can be approximated as:
\begin{equation}
    \nabla_{\rvx_\tau}\log p(\rvx_\tau) \approx \rvs_{\theta}(\rvx_\tau, \tau) = \frac{D_{\theta}(\rvx_\tau, \tau) - \rvx_\tau}{\sigma_\tau^2}.
\label{eq:score}
\end{equation}

\subsection{Decoupled Annealing Posterior Sampling}
The \ac{ODE} (\ref{eq:ode}) serves as the mechanism for unconditional sample generation from the prior data distribution $p(\rvx)$. In case a degraded observation of $\rvx$ is available, i.e. $\rvy=\gA(\rvx)+\rvv$ with $\gA$ being a degradation operator and $\rvv \sim \gN(\mathbf{0}, \sigma_\rvv^2\rmI)$, an estimate of $\rvx$ can be obtained by sampling from the posterior $p(\rvx\mid\rvy)$.

To pursue this, the \ac{DAPS} \cite{daps} framework proposed a sampling mechanism in which two intermediate states $\rvx_{\tau}, \rvx_{\tau+\Delta\tau}$ are independent when conditioned on $\rvx_0$. It was proven in \cite{daps} that for any $\tau_1, \tau_2$, if $\rvx_{\tau_1}$ is sampled from $p(\rvx_{\tau_1}\mid\rvy)$, then the following yields a sample from $p(\rvx_{\tau_2}\mid\rvy)$:
\begin{equation}
    \rvx_{\tau_2} \sim \E_{\rvx_0\sim p(\rvx_0\mid\rvx_{\tau_1},\rvy)}
    \left[\gN(\rvx_0, \sigma_{\tau_2}^2 \rmI)\right].
    \label{eq:daps}
\end{equation}
This translates into a dual-stage iterative strategy. The first step is sampling $\rvx_{0\mid\rvy}\sim p(\rvx_0\mid\rvx_{\tau_1}, \rvy)$. {The second step is sampling $\rvx_{\tau_2}\sim \gN(\rvx_{0\mid\rvy}, \sigma_{\tau_2}^2\rmI)$, i.e., the sample from the first step serves as the mean of the Gaussian in the second step. Note that for a sufficiently large $T_{\max}$, $p(\rvx_{T_{\max}}\mid\rvy)\approx p(\rvx_{T_{\max}})\approx \gN(\mathbf{0}, \sigma_{T_{\max}}^2\rmI)$.} Therefore, to sample from $p(\rvx\mid\rvy)$, first an annealed diffusion noise schedule for $\sigma_\tau$ with $N_A$ steps is determined between $T_{\max}$ and $T_{\min}$. Then, samples are iteratively drawn from $p(\rvx_{\tau}\mid\rvy)$ by following the two-stage procedure, starting from $\tau=T_{\max}$ until $\tau=T_{\min}$ is reached.

Sampling from $p(\rvx_0\mid\rvx_{\tau}, \rvy)$ is carried out by using methods for sampling from an unnormalised distribution. For example, Langevin dynamics \cite{langevin} is an iterative method, whose $j$-th update step is
\begin{align}
    \rvx_0^{(j+1)} = \rvx_0^{(j)} + \eta \nabla_{\rvx_0^{(j)}}\log p(\rvx_0^{(j)}\mid\rvx_{\tau}, \rvy) 
    + \sqrt{2 \eta}\,\boldsymbol{\epsilon}_j
    \label{eq:langevin1}
\end{align}
with $\eta > 0$ being the step size and $\boldsymbol{\epsilon}_j\sim \gN(\mathbf{0}, \rmI)$. By applying Bayes' Theorem, 
\begin{equation}
p(\rvx_0\mid\rvx_{\tau},\rvy) \propto p(\rvx_0\mid\rvx_{\tau})p(\rvy\mid\rvx_0).
\end{equation}
Additionally, \ac{DAPS} models $p(\rvx_0^{(j)}\mid\rvx_{\tau})$ as 
\begin{equation}
    p(\rvx_0 \mid \rvx_\tau) \approx \gN\bigl(\rvx_0; \hat{\rvx}_0(\rvx_\tau), r_\tau^2 \rmI\bigr)
\label{eq:prior}
\end{equation}
with $r_\tau$ chosen heuristically. $\hat{\rvx}_0(\rvx_\tau)$ is obtained by solving the ODE in (\ref{eq:ode}) with $N_{\text{ODE}}$ steps, unconditioned on $\rvy$. 
Consequently, the Langevin update rule can be rewritten as
\begin{align}
    \rvx_0^{(j+1)} = \rvx_0^{(j)} 
    &- \eta \nabla_{\rvx_0^{(j)}}\frac{\|\rvx_0^{(j)}-\hat{\rvx}_0(\rvx_\tau)\|^2}{2r_\tau^2} \nonumber\\
    &- \eta \nabla_{\rvx_0^{(j)}}\frac{\|\gA(\rvx_0^{(j)})-\rvy\|^2}{2\sigma_{\rvv}^2} 
    + \sqrt{2 \eta}\,\boldsymbol{\epsilon}_j.
    \label{eq:langevin2}
\end{align}
In practice, $\sigma_{\rvv}$ in the denominator is replaced with a user-controlled value $\alpha$ for better empirical performance. 
By repeating the Langevin iterations a sufficiently large number of times, $N_{\text{MC}}$, the distribution of $\rvx_0^{(N_{\text{MC}})}$ will be approximately $p(\rvx_0\mid\rvx_{\tau}, \rvy)$.

\section{Problem Formulation}
\label{sec:prob_form}
We consider a single-microphone time-domain observation $\rvy\in\sR^d$, where $d$ is the number of samples. Here $\rvy$ is a mixture signal formed by summing multiple independent signal components. Specifically, $\rvy$ consists of $K$ clean speech signals $\{\rvx^i\}_{i=1}^K\in\sR^d$ together with two additive noise terms of different type $\rvn,\rvz\in\sR^d$, yielding
\begin{equation}
    \rvy = \sum_{i=1}^{K} \rvx^i + \rvn + \rvz.
\label{eq:meas}
\end{equation}
All components are assumed to be statistically independent. The noise signal $\rvn$ models structured environmental interference, e.g., background street noises, while $\rvz$ denotes an auxiliary white Gaussian perturbation with vanishingly small variance $\sigma_\rvz^2$. This latter term is introduced to ensure that the resulting distribution has support throughout the signal space, thereby simplifying the mathematical treatment. Throughout the paper, we replace the set notation $\{\rvx_i\}_{i=1}^K$ with the concatenated vector $\rvc\in\sR^{Kd}$ for brevity. {Note that the degradation operator in our case is thus $\gA(\rvc,\rvn)=\sum_{i=1}^{K}\rvx^i+\rvn$, with $\rvz$ playing the role of $\rvv$.}

\section{Method}
The goal of our proposed method, SSNAPS, is to recover the concatenated speech signal $\rvc$ and the ambient noise $\rvn$ by sampling from the posterior $p(\rvc, \rvn\mid\rvy)$. To this end, in this section, we reformulate \ac{DAPS} to address our scenario, enabling it to estimate multiple statistically independent signals from two distinct distributions: clean speech and background noise. For the sake of simplicity in formulation, we assume that all speakers have visual cues in the form of lip videos that match their speech recordings. Later on, we present an extension to handle unseen speakers.

We start by defining $\rvh\in\sR^{(K+1)d}$ as the concatenation of $\rvc$ and $\rvn$. Hence, $p(\rvh\mid\rvy)=p(\rvx_1,...,\rvx_K, \rvn\mid\rvy)$, and the proof provided in \cite{daps} to justify (\ref{eq:daps}) holds in our case as well. Consequently, the decoupled sampling mechanism presented by \ac{DAPS} can be adopted by replacing $\rvx$ with $\rvh$ in (\ref{eq:daps}). Notably, each SSNAPS iteration comprises two steps: (1) sample $\rvh_{0\mid\rvy}\sim p(\rvh_0\mid\rvh_{\tau+\Delta\tau},\rvy)$, (2) sample $\rvh_{\tau}\sim \gN(\rvh_{0\mid\rvy}, \sigma_{\tau}^2\rmI)$. For conciseness, $\Delta\tau$ will be omitted henceforth when it is unnecessary for the clarity of the exposition.

We now focus on the first step of sampling from $p(\rvh_0\mid\rvh_{\tau+\Delta\tau},\rvy)$. Our preliminary evaluations showed that sampling with Langevin dynamics performed comparably to more sophisticated techniques, such as Hamiltonian Monte Carlo \cite{hmc}, and was therefore preferred for its relative simplicity.
Now, applying Bayes' Theorem to $p(\rvh_0\mid\rvh_{\tau},\rvy)$, and using the independence of all sources, yields
\begin{align}
    p(\rvh_0\mid&\rvh_{\tau},\rvy) \propto p(\rvy\mid\rvh_0)p(\rvh_0\mid\rvh_{\tau}) \nonumber
    \\ 
    & =p(\rvy \mid \rvc_0, \rvn_0) p(\rvn_0 \mid \rvn_\tau) \prod_{i=1}^K p(\rvx_0^i \mid \rvx_\tau^i).
    \label{eq:posterior}
\end{align}
A detailed derivation is provided in Appendix \ref{app:ssnaps}. 

Since all the sources are independent and only coupled via $\rvy$, the update step for the Langevin dynamics of (\ref{eq:posterior}) is equivalent to the following $K+1$ update rules:
\begin{subequations}
\begin{align}
    \rvx_0^{i^{(j+1)}} &= \rvx_0^{i^{(j)}} 
    + \eta \nabla_{\rvx_0^{i^{(j)}}} \log p\bigl(\rvx_0^{i^{(j)}} \mid \rvx_\tau^i\bigr) \nonumber\\
    &+ \eta \nabla_{\rvx_0^{i^{(j)}}} \log p\bigl(\rvy \mid \rvc_0^{(j)},\rvn_0^{(j)}\bigr)
    + \sqrt{2 \eta}\,\boldsymbol{\epsilon}_j^{\rvx^i}
    \label{eq:langevin_sp}
    \\
    \rvn_0^{{(j+1)}} &= \rvn_0^{{(j)}} 
    + \eta \nabla_{\rvn_0^{{(j)}}} \log p\bigl(\rvn_0^{{(j)}} \mid \rvn_\tau\bigr) \nonumber\\
    &+ \eta \nabla_{\rvn_0^{(j)}} \log p\bigl(\mathbf{y} \mid \rvc_0^{(j)},\rvn_0^{(j)}\bigr)
    + \sqrt{2 \eta}\,\boldsymbol{\epsilon}_j^{\rvn}
    \label{eq:langevin_np}
\end{align}
\end{subequations}
for $i=1\ldots,K$. $\boldsymbol{\epsilon}_j^{\rvx^i},\boldsymbol{\epsilon}_j^{\rvn} \sim \gN(\mathbf{0}, \rmI)$ are signal specific Gaussian noise terms. More details on the derivations may be found in Appendix \ref{app:ssnaps}.

In line with (\ref{eq:prior}), we can write for $i=1,\ldots,K$:
\begin{subequations}
\begin{align}
    p(\rvx_0^i \mid \rvx^i_\tau) &\approx \gN\bigl(\rvx_0^i; \hat{\rvx}_0^i(\rvx^i_\tau), r_\tau^2 \rmI\bigr) \\
    p(\rvn_0 \mid \rvn_\tau) &\approx \gN\bigl(\rvn_0; \hat{\rvn}_0(\rvn_\tau), r_\tau^2 \rmI\bigr).
\end{align}
\end{subequations}
Similarly to \ac{DAPS}, \{$\hat{\rvx}_0^i(\rvx^i_\tau)\}_{i=1}^K$ and $\hat{\rvn}_0(\rvn_\tau)$ are obtained by sampling from the following K+1 \acp{ODE} using $N_{\text{ODE}}$ steps:
\begin{subequations}
\begin{align}
    &\mathrm{d}\rvx^i_\tau = -\rvs_{\theta}(\rvx^i_\tau, \tau, V_i)\cdot \tau \mathrm{d}\tau 
    \label{eq:ode_speech}
    \\
    &\mathrm{d}\rvn_\tau = -\rvs_{\phi}(\rvn_\tau, \tau)\cdot \tau \mathrm{d}\tau.
\label{eq:ode_noise}
\end{align}
\end{subequations}
$\rvs_{\phi}(\rvn_\tau, \tau)$ is the score function of the noise prior, which utilises the diffusion denoiser $G_{\phi}(\rvn_\tau,\tau)$ in accordance with (\ref{eq:score}). The speech score computation is implemented with the audio-visual score function $\rvs_{\theta}(\rvx^i_\tau, \tau, V_i)$. $V_i$ are the visual features derived from the lip video associated with the $i$th speaker, which help enhance the accuracy of the score estimation. $\rvs_{\theta}(\rvx^i_\tau, \tau, V_i)$ deploys $F_\theta(\rvx^i_\tau,\tau,V_i)$ as the denoiser in conjunction with the classifier-free guidance paradigm \cite{ho2021classifierfree}. Note that all speech signals share the same speech score function.

The likelihood terms in (\ref{eq:langevin_sp}), (\ref{eq:langevin_np}) are proportional to the mixture reconstruction error:
\begin{subequations}
    \begin{align}
    \nabla_{\rvx_0^{i^{(j)}}} \log p\bigl(\mathbf{y} \mid \rvc_0^{(j)},\rvn_0^{(j)}\bigr) &= -\nabla_{\rvx_0^{i^{(j)}}} \frac{\gL_{\text{rec}}}{2\sigma_\rvz^2} \\
    \nabla_{\rvn_0^{(j)}} \log p\bigl(\mathbf{y} \mid \rvc_0^{(j)},\rvn_0^{(j)}\bigr) &= -\nabla_{\rvn_0^{(j)}} \frac{\gL_{\text{rec}}}{2\sigma_\rvz^2}.
    \end{align}    
\end{subequations}
Although by rigorous calculus $\gL_{\text{rec}}$ is a time domain mixture reconstruction loss, we follow \cite{buddy,davssnm} by defining $\gL_{\text{rec}}$ in the frequency domain to achieve more accurate gradients:
\begin{align}
\gL_{\text{rec}} =
\left\|S(\rvy) - S\left(\sum_{i=1}^K \rvx_0^{i^{(j)}} + \rvn_0^{(j)})\right)\right\|_2^2,
\end{align}
where $S(\rvy)=|\textrm{STFT}(\rvy)|^{\frac{2}{3}}\exp{j\angle \textrm{STFT}(\rvy)}$, and $\textrm{STFT}(\cdot)$ is the \ac{STFT}.

In conclusion, the $j$-th Langevin update in SSNAPS is:
\begin{subequations}
\begin{align}
    \rvx_0^{i^{(j+1)}} = \rvx_0^{i^{(j)}} 
    &- \eta \nabla_{\rvx_0^{i^{(j)}}} \frac{\|\rvx_0^{i^{(j)}}
    -\hat{\rvx}_0^i(\rvx_\tau^i)\|^2}{2r_\tau^2}  \nonumber\\
    &- \eta \nabla_{\rvx_0^{i^{(j)}}}\frac{\gL_{\text{rec}}}{2\sigma_\rvz^2}
    + \sqrt{2 \eta}\,\boldsymbol{\epsilon}_j^{\rvx^i} 
    \label{eq:langevin_s}\\
    \rvn_0^{{(j+1)}} = \rvn_0^{{(j)}} 
    &- \eta \nabla_{\rvn_0^{(j)}}\frac{\|\rvn_0^{(j)}-\hat{\rvn}_0(\rvn_\tau)\|^2}{2r_\tau^2}  \nonumber\\
    &- \eta \nabla_{\rvn_0^{(j)}}\frac{\gL_{\text{rec}}}{2\sigma_\rvz^2}
    + \sqrt{2 \eta}\,\boldsymbol{\epsilon}_j^{\rvn}.
    \label{eq:langevin_n}
\end{align}
\end{subequations}
Like DAPS, SSNAPS also replaces $\sigma_\rvz$ in the denominator with a user-defined parameter $\alpha$.
A summary of SSNAPS is provided in Algorithm \ref{alg:ssnaps}. 

As the experimental study in Section \ref{sec:results} demonstrates, $\gL_{\text{rec}}$ suffices to make SSNAPS perform well, even without introducing additional loss terms which limit potential crosstalk between the separated components. This can be attributed to the strong guidance signal induced by the visual features associated with the speech signals. 

\begin{algorithm*}[tb]
  \caption{SSNAPS}
  \begin{algorithmic}
    \STATE {\bfseries Input:} mixture $\rvy$, score models $\rvs_{\theta}, \rvs_{\phi}$, diffusion noise schedule $\sigma_\tau$, $(\tau_m)_{m\in\{0,\ldots,N_A\}}$
    \FOR{$m=N_A,N_A-1,\ldots1$}
        \STATE Compute $\rvn_0^{(0)}=\hat{\rvn}_0(\rvn_{\tau_m})$ by solving $\mathrm{d}\rvn_\tau = -\rvs_{\phi}(\rvn_\tau, \tau)\cdot \tau \mathrm{d}\tau$ initialised with $\rvn_{\tau_m}$ 
        \hfill $\triangleright$ $N_{\text{ODE}}$ steps
        \FOR{$i=1,\ldots,K$}
            \STATE Compute $\rvx_0^{i^{(0)}}=\hat{\rvx}_0^i(\rvx_{\tau_m}^i)$ by solving $\mathrm{d}\rvx^i_\tau = -\rvs_{\theta}(\rvx^i_\tau, \tau, V_i)\cdot \tau \mathrm{d}\tau$ initialised with $\rvx_{\tau_m}^i$ \hfill $\triangleright$ $N_{\text{ODE}}$ steps
            \ENDFOR
        \FOR{$j=0,\ldots,N_{\text{MC}}-1$}
            \STATE \textit{Langevin dynamics:}
            \STATE $\gL_{\text{rec}} =
                    \left\|S(\rvy) - S\left(\sum_{i=1}^K \rvx_0^{i^{(j)}} + \rvn_0^{(j)})\right)\right\|_2^2$
            \STATE Draw $\boldsymbol{\epsilon}_j^{\rvn}\sim\gN(\mathbf{0}, \rmI)$
            \STATE $\rvn_0^{{(j+1)}} = \rvn_0^{{(j)}} 
            - \eta \nabla_{\rvn_0^{(j)}}\frac{\|\rvn_0^{(j)}-\hat{\rvn}_0(\rvn_\tau)\|^2}{\sigma_\tau^2}  
            - \eta \nabla_{\rvn_0^{(j)}}\frac{\gL_{\text{rec}}}{\alpha^2}
            + \sqrt{2 \eta}\,\boldsymbol{\epsilon}_j^{\rvn}$ \hfill $\triangleright$ Noise Langevin Update
            \FOR{$i=1,\ldots,K$}
                \STATE Draw $\boldsymbol{\epsilon}_j^{\rvx^i}\sim\gN(\mathbf{0}, \rmI)$
                \STATE $\rvx_0^{i^{(j+1)}} = \rvx_0^{i^{(j)}} 
                - \eta \nabla_{\rvx_0^{i^{(j)}}}\frac{\|\rvx_0^{i^{(j)}}
                -\hat{\rvx}_0^i(\rvx_\tau^i)\|^2}{\sigma_\tau^2}  
                - \eta \nabla_{\rvx_0^{i^{(j)}}}\frac{\gL_{\text{rec}}}{\alpha^2}
                + \sqrt{2 \eta}\,\boldsymbol{\epsilon}_j^{\rvx^i}$ \hfill $\triangleright$ Speech Langevin Update
            \ENDFOR
        \ENDFOR
        \STATE Sample $\rvn_{\tau_{m-1}}\sim\gN(\rvn_0^{(N_{\text{MC}})}, \sigma_{\tau_{m-1}}^2\rmI)$
        \FOR{$i=1,\ldots,K$}
            \STATE Sample $\rvx_{\tau_{m-1}}^i\sim\gN(\rvx_0^{i^{(N_{\text{MC}})}}, \sigma_{\tau_{m-1}}^2\rmI)$
        \ENDFOR
    \ENDFOR
    \STATE \textbf{return} $\rvx_0^1,\ldots,\rvx_0^K,\rvn_0$
  \end{algorithmic}
\label{alg:ssnaps}
\end{algorithm*}

\section{SSNAPS for Off-Screen Speech Separation}
The video for each speaker guides the inverse sampling process to match the speaker's lip movements. It helps discern, e.g., between speakers with similar pitch, and between speech and noise with overlapping spectral profiles. However, in many real-world cases, not all speech recordings have corresponding lip videos. 

In the absence of visual guidance, without special care, separation techniques may be prone to signal leakage between estimated sources, also known as crosstalk. This is manifested as one source being audible in the recovered signal of another source. The primary cause of this crosstalk is typically the difficulty of the separation model in distinguishing between acoustically similar sources, for example, in separating same-sex speech. Although off-screen speech separation is common in real-world settings, this field remains underexplored. While a few supervised paradigms have been proposed \cite{ravss,offscreen}, to the best of our knowledge, we are the first to address this challenge in the unsupervised regime.

Here we assume that among $K$ speakers, only $K-1$ have visual cues. The $K$-th speaker is arbitrarily defined as the one lacking visual guidance. We further assume that $K>1$, i.e., our focus is on joint speech separation and enhancement. Arguably, in many real-world acoustic scenes there are only two or three speakers. Consequently, it is plausible to assume that only one speaker (for $K=2$) or two speakers (for $K=3$) have visual cues.
Since SSNAPS is an unsupervised technique, only the case of a single video-less speaker is considered. {Otherwise, with several such speakers who share the same visually unguided speech prior, discerning between speakers with similar pitch becomes difficult. Since the likelihood only depends on the sum of the sources, this can give rise to speech attribution ambiguity and source permutation. This setback cannot be resolved without major methodological changes, which are left for future work.}

Now, recall that the speech diffusion denoiser $F_\theta$ is trained under the classifier-free guidance methodology. Thus, the score function $\rvs_\theta$ can be estimated with $F_\theta(\rvx^i_\tau,\tau,\varnothing)$ instead of $F_\theta(\rvx^i_\tau,\tau,V_i)$, where $\varnothing$ is a null-token representing the state of a missing lip video. It is, however, still necessary to adapt SSNAPS to minimise potential crosstalk. 


In our experiments on off-screen speaker separation, we found that crosstalk between the visually guided and unguided speech sources was the main cause of degradation. We thus propose to address the off-screen speech separation setting by introducing a loss term that is integrated into the Langevin update equations (\ref{eq:langevin_s}) and (\ref{eq:langevin_n}). It is designed to minimise signal leakage between the estimated visually guided and unguided speech components, imposed via the following loss function:
\begin{equation}
    \gL_{\textrm{ct-ss}} = \sum_{i=1}^{K-1}\mathrm{sim} \left(|S(\rvx^i)|, \textrm{sg}(|S(\rvx^K)|)\right).
\end{equation}
Here $\mathrm{sim}(A,B)$ computes the cosine similarity between $A$ and $B$. The $\mathrm{sg}(\cdot)$ function renders its argument a constant with respect to the gradient operation applied to the loss function in the Langevin dynamics equations.  Crucially, only speech signals with visual cues are allowed to contribute to the minimisation of the loss, as their visual guidance is expected to yield better gradients.


Finally, the Langevin update step for the off-screen speech separation setting is identical to (\ref{eq:langevin_s}), (\ref{eq:langevin_n}), but with $\gL_{\textrm{os}}$ instead of $\gL_{\text{rec}}$, such that
\begin{equation}
\gL_{\textrm{os}} = \gL_{\text{rec}} 
+ g_{\textrm{ct-ss}}\gL_{\textrm{ct-ss}}.
\end{equation}
$g_{\textrm{ct-ss}}$ is a scaling factor. Additionally, $\gL_{\textrm{ct-ss}}$ is applied only for $\sigma<\sigma_{\textrm{os}}$, to allow the estimated sources to be in the vicinity of their respective manifolds first. Otherwise, for $\sigma\geq\sigma_{\textrm{os}}$, only $\gL_{\textrm{rec}}$ is utilised. A summary of SSNAPS for off-screen speech separation is provided in Algorithm \ref{alg:ssnaps_os}.

\begin{algorithm*}[tb]
  \caption{SSNAPS for off-screen speech separation}
  \begin{algorithmic}
    \STATE {\bfseries Input:} mixture $\rvy$, $\sigma_{\textrm{os}}$, $g_{\textrm{ct-ss}}$, score models $\rvs_{\theta}, \rvs_{\phi}$, diffusion noise schedule $\sigma_\tau$, $(\tau_m)_{m\in\{0,\ldots,N_A\}}$
    \FOR{$m=N_A,N_A-1,\ldots1$}
        \STATE Compute $\rvn_0^{(0)}=\hat{\rvn}_0(\rvn_{\tau_m})$ by solving $d\rvn_\tau = -\rvs_{\phi}(\rvn_\tau, \tau)\cdot \tau d\tau$ initialised with $\rvn_{\tau_m}$ 
        \hfill $\triangleright$ $N_{\text{ODE}}$ steps
        \FOR{$i=1,\ldots,K$}
            \STATE Compute $\rvx_0^{i^{(0)}}=\hat{\rvx}_0^i(\rvx_{\tau_m}^i)$ by solving $d\rvx^i_\tau = -\rvs_{\theta}(\rvx^i_\tau, \tau, V_i)\cdot \tau d\tau$ initialised with $\rvx_{\tau_m}^i$ \hfill $\triangleright$ $N_{\text{ODE}}$ steps
            \ENDFOR
        \FOR{$j=0,\ldots,N_{\text{MC}}-1$}
            \STATE \textit{Langevin dynamics:}
            \STATE $\gL_{\text{rec}} =
                    \left\|S(\rvy) - S\left(\sum_{i=1}^K \rvx_0^{i^{(j)}} + \rvn_0^{(j)})\right)\right\|_2^2$
            \STATE Draw $\boldsymbol{\epsilon}_j^{\rvn}\sim\gN(\mathbf{0}, \rmI)$
            \STATE $\rvn_0^{{(j+1)}} = \rvn_0^{{(j)}} 
            - \eta \nabla_{\rvn_0^{(j)}}\frac{\|\rvn_0^{(j)}-\hat{\rvn}_0(\rvn_\tau)\|^2}{\sigma_\tau^2}  
            - \eta \nabla_{\rvn_0^{(j)}}\frac{\gL_{\text{rec}}}{\alpha^2}
            + \sqrt{2 \eta}\,\boldsymbol{\epsilon}_j^{\rvn}$ \hfill $\triangleright$ Noise Langevin Update
            \FOR{$i=1,\ldots,K$}
                \STATE Draw $\boldsymbol{\epsilon}_j^{\rvx^i}\sim\gN(\mathbf{0}, \rmI)$
                \IF{$\sigma_{\tau_m} < \sigma_{\textrm{os}}$ {\bfseries and} $i\neq K$}
                \STATE $\gL_{\textrm{ct-ss}} = \mathrm{sim} \left(|S(\rvx^i)|, \textrm{sg}(|S(\rvx^K)|)\right)$
                \STATE $\gL_{\text{os}} = \gL_{\text{rec}} + g_{\textrm{ct-ss}}\gL_{\textrm{ct-ss}}$
                \ELSE
                \STATE $\gL_{\text{os}} = \gL_{\text{rec}}$
                \ENDIF
                \STATE $\rvx_0^{i^{(j+1)}} = \rvx_0^{i^{(j)}} 
                - \eta \nabla_{\rvx_0^{i^{(j)}}}\frac{\|\rvx_0^{i^{(j)}}
                -\hat{\rvx}_0^i(\rvx_\tau^i)\|^2}{\sigma_\tau^2}  
                - \eta \nabla_{\rvx_0^{i^{(j)}}}\frac{\gL_{\text{os}}}{\alpha^2}
                + \sqrt{2 \eta}\,\boldsymbol{\epsilon}_j^{\rvx^i}$ \hfill $\triangleright$ Speech Langevin Update
            \ENDFOR
        \ENDFOR
        \STATE Sample $\rvn_{\tau_{m-1}}\sim\gN(\rvn_0^{(N_{\text{MC}})}, \sigma_{\tau_{m-1}}^2\rmI)$
        \FOR{$i=1,\ldots,K$}
            \STATE Sample $\rvx_{\tau_{m-1}}^i\sim\gN(\rvx_0^{i^{(N_{\text{MC}})}}, \sigma_{\tau_{m-1}}^2\rmI)$
        \ENDFOR
    \ENDFOR
    \STATE \textbf{return} $\rvx_0^1,\ldots,\rvx_0^K,\rvn_0$
  \end{algorithmic}
\label{alg:ssnaps_os}
\end{algorithm*}

\section{Experimental Setup}
In this section, we present the experimental settings used to evaluate the proposed method. We describe the datasets used for evaluations and key implementation details. 

\subsection{Setup}
\subsubsection{Data} 
The clean speech dataset used for training and evaluation was VoxCeleb2 \cite{vox2}. The English-only version \cite{avhubert} of the audio-visual dataset was used to align with the dataset on which the visual encoder was trained. This amounted to roughly 628,000 videos, equivalent to 1,759 hours. 2,000 videos were reserved for testing. The audio and video have sampling rates of 16 kHz and 25 Hz, respectively. For SSNAPS, lip-region videos are extracted from full-face videos by following the prescription in \cite{braven}. 

For ambient noise recordings, two datasets are used. The first dataset is DNS \cite{dns}, which is used for both training and testing. It is a subset of AudioSet \cite{audioset} with additional recordings from Freesound.\footnote{\url{https://freesound.org/}}
Overall, it consists of about 65,000 clips from 150 noise classes, 181 hours of recordings in total. We use the train/validation/test splits defined by \cite{davssnm}, resulting in 64,000 noise signals for training, 100 for validation, and 900 for evaluation. 

The second noise dataset was DCASE 2020 Challenge \cite{dcase}, which was used only for evaluation. It contains data recorded in ten European cities and nine devices. There are ten noise classes, and all recordings have a fixed length of 10~s, sampled at 44.1~kHz, which we downsampled to 16~kHz. Only the test split was used, 19 hours in total. 

\subsubsection{Networks architecture} The speech diffusion denoiser $F_\theta$ was trained on 4s-long clean speech signals from the English-only version of VoxCeleb2. The background noise diffusion denoiser $G_\phi$ was also trained on 4s signals, but of noise recordings drawn from DNS. They are both based on the lightweight U-Net \cite{unet} architecture NCSN++M proposed in \cite{storm}. The NCSN++M backbones of $F_\theta$ and $G_\phi$ are wrapped with STFT and inverse STFT, similarly to \cite{buddy}. 
All \ac{STFT} operations use a Hann window of 510 samples and a hop length of 160. This results in 256 frequency bins.

For $F_\theta$, the NCSN++M backbone is augmented with visual features integrated via FiLM layers \cite{film}, and was trained with classifier-free guidance. The lip video frames are encoded using the BRAVEn model \cite{braven} finetuned on visual speech recognition. Each lip video frame is encoded to a feature vector in $\sR^{1024}$. Altogether, the number of trainable parameters of $F_\theta$ and $G_\phi$ is 129.5M and 39.7M, respectively.

\subsubsection{Baselines}
SSNAPS is primarily compared against two top-tier audio-visual \emph{supervised} baselines, which we trained from scratch.  
The first competing supervised method is FlowAVSE \cite{flowavse}, an audio-visual speaker extraction method. Its architecture is based on cascaded discriminative and generative models. The generative model is based on flow matching, which aims to improve the signal quality of the extracted speech at the output of the discriminative network. FlowAVSE has 60.2M parameters.
The second supervised baseline is RAVSS \cite{ravss}, a recently proposed powerful audio-visual discriminative technique based on the transformer architecture \cite{attention} with 24.3M trainable parameters. Its design allows it to separate a variable number of speakers using a single model and to address scenarios with missing visual streams.
Both baselines outperform several previously proposed discriminative and generative audio-visual paradigms. Since RAVSS did not originally consider noisy mixtures, we treat the noise as an additional source lacking visual features.

We train FlowAVSE and RAVSS on speech-noise combinations from VoxCeleb2+DNS, where the \ac{SIR} and \ac{SNR} used to create the training mixtures matched the evaluation protocol described in the sequel. Importantly, during RAVSS training, visual streams are randomly discarded to make the model robust to missing visual features.
To demonstrate the performance improvement provided by SSNAPS relative to unsupervised techniques, we also compare it with DAVSS-NM \cite{davssnm} for two-speaker speech separation. It also utilises speech and environmental noise priors and estimates the individual sources using \ac{DPS} as the inverse sampler. {SSNAPS and DAVSS-NM use the same noise prior mode, with the same pretrained checkpoint, visual encoder and preprocessing pipeline.}

{In addition to off-screen experiments, to assess the contribution of the visual modality, we also consider audio-only enhancement of single-speaker VoxCeleb2+DNS mixtures. For SSNAPS, the visual conditioning of $F_\theta$ replaced by the null token $\varnothing$ at all sampling steps, which requires no retraining. SSNAPS is compared with SGMSE \cite{sgmse}, a supervised score-based method, and UDiffSE \cite{nmf1}, an unsupervised diffusion-based method that combines a clean-speech diffusion prior with an \ac{NMF} noise model. We train SGMSE on VoxCeleb2+DNS mixtures, and the clean-speech prior of UDiffSE is trained on VoxCeleb2.}

\subsubsection{Evaluation protocol}
All methods were evaluated on mixtures of 1, 2, and 3 speech signals, each with a single background noise recording. All mixtures were 4 seconds long, resulting in $d=64,000$. Each scenario comprised 500 mixtures. In the single-speaker case, for each mixture, the \ac{SNR} was randomly sampled from the range [-5, 10]~dB. In the two-speaker setup, the \ac{SIR} values were sampled from [-5,5]~dB, and the \ac{SNR} was randomly sampled from [-3,3]~dB, defined with respect to the low-power speaker. Finally, when three speakers are present, the \ac{SIR} was 0 dB, and the \ac{SNR} was set to 15~dB to avoid overly adverse conditions. 

To examine the ability of all methods to generalise, two scenarios were considered. In the first, matched scenario, all methods were trained and tested on mixtures from VoxCeleb2+DNS. In the unmatched regime, the \emph{test} noise dataset is DCASE instead of DNS. Moreover, for off-screen speaker separation, only 2- and 3-speaker mixtures are considered, using the same \ac{SIR} and \ac{SNR} ranges described above. In addition, the demo page\footnote{\label{fn:demo}\url{https://ssnaps2026.github.io/ssnaps2026/}} includes video samples which demonstrate that SSNAPS can successfully deal with real world videos, with recordings of various lengths (different values of $d$). In particular, we also include videos longer than 4 seconds.

We use both acoustic and semantic metrics to quantify the performance achieved by each method. The acoustic metrics are further categorised into reference-dependent and reference-free, with the reference being clean speech. The former includes \ac{PESQ} score, \ac{ESTOI}, and \ac{SI-SDR}, while for the latter \ac{NISQA} \cite{nisqa1,nisqa2} is deployed. Finally, to assess the semantic quality of the separated sources, \ac{WER} is used. We report the \ac{WER} using two leading \acp{ASR}:  NVIDIA’s FastConformer~\cite{rekesh2023fast} and Whisper~\cite{radford2022whisper}. Since the VoxCeleb2 does not provide ground-truth transcriptions, we use the transcriptions of the clean speech signals produced by the \acp{ASR} as pseudo ground truth when computing \ac{WER}. More details on the \acp{ASR} and the \ac{WER} computations can be found in Sec.~\ref{app:wer}.


\subsection{Implementation Details}
\label{app:imp}
\subsubsection{Annealing Scheduler} SSNAPS uses an annealed noise schedule with $N_A$ steps, where in each step posterior sampling is carried out via Langevin dynamics yielding $\hat{\rvh}_0$, followed by white noise perturbation $\rvh_{\tau}=\hat{\rvh}_0+\sigma_\tau\boldsymbol{\epsilon}$, $\boldsymbol{\epsilon}\sim\gN(\mathbf{0},\rmI)$. The annealing noise schedule is determined according to \cite{karras2022elucidating}:
\begin{equation}
    \tau_i = \left( \tau^{\frac{1}{\rho}} + \frac{i}{N_A-1} \left( \tau_{\min}^{\frac{1}{\rho}} - \tau^{\frac{1}{\rho}} \right) \right)^{\rho}
    \label{eq:placeholder_label}
\end{equation}
where $\rho=10$ is used in all of our experiments. We use the parameterisation $\sigma_\tau^A=\tau$. The last annealing step $\sigma^A_{T_{\min}}$ is set to 0.01 for all tasks. The values of $\sigma_{T_{\max}}^A$ and $N_A$ are task-specific and are provided in Table~\ref{tb:imp}.

\subsubsection{\ac{ODE} Solver} The \ac{ODE} was solved with $N_{\text{ODE}}=2$ Euler steps, where the parameterisation $\sigma_\tau^{\text{ODE}}=\tau$ \cite{karras2022elucidating}, and $\tau_{\min}=1e-5$ are used across all tasks. Note that $\tau_{\max}$ is determined by $\sigma_\tau^A$. The value of the classifier-free guidance parameter $\omega$ used for speech score computations is provided in Table~\ref{tb:imp} for each task.

\subsubsection{Langevin Dynamics}
Similarly to \cite{daps}, we adopt a step-dependent learning rate schedule. The learning rate of the $j$-th Langevin update is $\eta_j=\eta_0[\delta+j/{N_{\text{MC}}}(1-\delta)]$, where $\delta$ is a decay ratio. We use $\delta=0.01$ and $\eta_0=1e-6$ for all tasks. Additionally, the value of $\alpha$, which scales the likelihood term, is task-specific and provided in Table~\ref{tb:imp}.

\subsubsection{Off-Screen Speech Separation}
The hyperparameters used for the 2-speaker+noise and 3-speaker+noise off-screen speaker separation experiments are identical to those used in tests where all speakers have visual streams. For the 2-speaker+noise speech separation task, the values of $g_{\textrm{ct-ss}}$ and $\sigma_{\textrm{os}}$ are determined as 20 and 0.25, respectively, and for the 3-speaker+noise task, they are set to 5 and 0.14, respectively.

\begin{table}
  \caption{Values of hyperparameters used by SSNAPS for the different tasks.}
  \label{tb:imp}
  \begin{center}
    \begin{small}
      \begin{sc}
        \begin{tabular}{lccccc} 
        \toprule
        Task & $N_{A}$ & $N_{\text{MC}}$ & $\sigma^A_{T_{\max}}$ & $\alpha$ & $\omega$ \\
        \midrule
        1 speaker+noise  & 300 & 50 & 2 & 0.0005 & 0.8 \\
        2 speakers+noise & 300 & 100 & 4 & 0.001 & 0.8 \\
        3 speakers+noise & 400 & 100 & 3 & 0.001 & 0.5 \\
        \bottomrule
        \end{tabular}
      \end{sc}
    \end{small}
  \end{center}
  \vskip -0.1in
\end{table}

\subsection{\ac{WER} Computation Details}
\label{app:wer}
To assess the impact of our method when used as input to an \ac{ASR} system, we evaluate the \ac{WER} of the separated output signals. We employ two state-of-the-art \ac{ASR} models and report results for both. The first is NVIDIA’s FastConformer~\cite{rekesh2023fast} with a \ac{TDT} decoder~\cite{xu2023efficient}, referred to as Parakeet-TDT, using the 0.6B-v2 variant. The second is the latest Whisper model~\cite{radford2022whisper} (at the time of submission), specifically the Whisper-large-v3 variant. Both models represent strong, publicly available \ac{ASR} systems and achieve low \ac{WER} on standard benchmark datasets.

\subsubsection{Corpus-Level {WER}}
In multi-speaker mixtures, \acf{ASR} often produces a large number of errors due to overlapping speech, resulting in very high \ac{WER} values prior to applying any separation algorithm. By design, the \ac{WER} score can exceed 100\% when the total number of substitutions, deletions, and insertions is greater than the number of words in the reference transcription.

Recall that the \ac{WER} for a single utterance \(i\) is defined as
\begin{equation}
\mathrm{WER}_i = \frac{S_i + D_i + I_i}{N_i},
\end{equation}
where \(S_i\), \(D_i\), and \(I_i\) denote the number of substitutions, deletions, and insertions, respectively, computed for a single utterance (indexed $i$), and \(N_i\) is the number of words in the corresponding reference transcription.

When computing \ac{WER} over large datasets containing utterances of varying lengths, this definition can lead to high variance, as longer utterances contribute disproportionately to the overall score. Consequently, instead of computing \ac{WER} independently for each utterance and aggregating the scores using the mean and standard deviation, we compute a corpus-level \ac{WER}. The corpus-level \ac{WER} is defined as
\begin{equation}
\mathrm{WER}_{\mathrm{corpus}} = \frac{\sum_{i=1}^{M} \left(S_i + D_i + I_i\right)}{\sum_{i=1}^{M} N_i},
\end{equation}
where \(M\) denotes the number of utterances in the corpus.

This formulation ensures that longer utterances contribute proportionally to the overall score, resulting in a more stable evaluation across datasets with heterogeneous utterance lengths. Corpus-level \ac{WER} is the standard evaluation protocol in \ac{ASR}, and unless stated otherwise, all \ac{WER} results reported in this work correspond to corpus-level \ac{WER}.

\begin{table*}[t]
  \caption{Speech separation performance for VoxCeleb2+DNS 1, 2 and 3 speakers+noise mixtures with standard deviations, except for \ac{WER} which was computed on the corpus. Best results are in bold, and runners-up are underlined. \#Spk denotes the number of speakers in the mixture. Three different models were trained for FlowAVSE, corresponding to the number of speakers. They are all evaluated for each mixture type. Unsup. indicates if the method is unsupervised. The four WER columns are denoted by two letters, where the first indicates the \ac{ASR} used to produce the pseudo ground-truth transcription from the clean speech and the second the \ac{ASR} applied to the evaluated signal (N: NVIDIA Parakeet-TDT, W: Whisper-large-v3). 
  }
  \label{tb:dns1}
  \begin{center}
      \begin{sc}
        \begin{tabular}{lccccccrrrr} 
            \toprule
            \multirow{2}{*}{Method} & \multirow{2}{*}{\#Spk} & \multirow{2}{*}{Unsup.} & \multirow{2}{*}{SISDR$\uparrow$} & \multirow{2}{*}{PESQ$\uparrow$} & \multirow{2}{*}{ESTOI$\uparrow$} & \multirow{2}{*}{NISQA$\uparrow$} & \multicolumn{4}{c}{WER$\downarrow$} \\
            \cmidrule(lr){8-11}
            & & & & & & & N-N & W-W & {N-W} & {W-N} \\
            \midrule
            Mixture & 1 & - & 2.61$\pm$4.28 & 1.95$\pm$0.55 & 0.61$\pm$0.17 & 1.36$\pm$0.72 & 13.3 & 14.1 & {19.2} & {18.0} \\
            \hdashline
            FlowAVSE \cite{flowavse} & & & & & & & & & & \\
            \hspace{2mm}1 spk & 1 & \xmark & \textbf{14.09$\pm$4.54} & \textbf{3.01$\pm$0.75} & \textbf{0.82$\pm$0.12} & \textbf{3.84$\pm$0.84} & \underline{12.5} & \textbf{11.5} & {\underline{17.4}} & {\textbf{15.7}} \\
            \hspace{2mm}2 spk & 1 & \xmark & 10.40$\pm$4.21 & 2.67$\pm$0.69 & 0.77$\pm$0.12 & \textbf{3.84$\pm$0.70} & 14.9 & 18.4 & {18.8} & {21.1} \\
            \hspace{2mm}3 spk & 1 & \xmark & 0.72$\pm$3.01 & 1.98$\pm$0.61 & 0.67$\pm$0.11 & \underline{3.15$\pm$0.70} & 24.1 & 23.4 & {27.6} & {26.0} \\
            RAVSS \cite{ravss} & 1 & \xmark & \underline{12.26$\pm$4.26} & \underline{2.94$\pm$0.56} & \underline{0.78$\pm$0.12} & 2.45$\pm$0.58 & 13.2 & 13.5 & {17.9} & {17.7} \\
            SSNAPS (ours) & 1 & \checkmark & 11.83$\pm$4.39 & 2.73$\pm$0.63 & 0.76$\pm$0.13 & 2.61$\pm$0.65 & \textbf{11.3} & \underline{11.8} & {\textbf{15.7}} & {\underline{15.9}} \\
            \midrule
            Mixture & 2 & - & -2.66$\pm$3.03 & 1.46$\pm$0.23 & 0.38$\pm$0.11 & 1.11$\pm$0.53 & 115.1 & 113.3 & {113.9} & {114.4} \\
            \hdashline
            FlowAVSE \cite{flowavse} & & & & & & & & & & \\
            \hspace{2mm}1 spk & 2 & \xmark & -1.96$\pm$8.54 & 1.57$\pm$0.43 & 0.43$\pm$0.20 & 3.02$\pm$0.87 & 83.5 & 71.2 & {83.2} & {72.1} \\
            \hspace{2mm}2 spk & 2 & \xmark & \underline{7.82$\pm$4.87} & 2.15$\pm$0.51 & \underline{0.65$\pm$0.14} & \textbf{3.48$\pm$0.75} & 30.2 & 28.9 & {32.3} & {30.2} \\
            \hspace{2mm}3 spk & 2 & \xmark & 3.85$\pm$4.76 & 1.96$\pm$0.44 & 0.61$\pm$0.13 & \underline{3.30$\pm$0.74} & 38.1 & 37.7 & {40.1} & {37.8} \\
            RAVSS \cite{ravss} & 2 & \xmark & \textbf{8.64$\pm$2.92} & \textbf{2.40$\pm$0.41} & \underline{0.65$\pm$0.12} & 1.99$\pm$0.47 & \underline{24.1} & \underline{23.4} & {\underline{26.9}} & {\underline{25.6}} \\
            DAVSS-NM \cite{davssnm} & 2 & \checkmark & 5.06$\pm$4.44 & 2.06$\pm$0.45 & 0.61$\pm$0.14 & 3.10$\pm$0.87 & 27.5 & 26.8 & {29.6} & {28.7} \\
            SSNAPS (ours) & 2 & \checkmark & 7.45$\pm$3.75 & \underline{2.25$\pm$0.48} & \textbf{0.66$\pm$0.12} & 2.44$\pm$0.65 & \textbf{19.2} & \textbf{18.3} & {\textbf{22.4}} & {\textbf{21.2}} \\
            \midrule
            Mixture & 3 & - & -3.08$\pm$0.15 & 1.43$\pm$0.14 & 0.34$\pm$0.07 & 0.87$\pm$0.39 & 171.0 & 205.8 & {168.5} & {194.5} \\
            \hdashline
            FlowAVSE \cite{flowavse} & & & & & & & & & & \\
            \hspace{2mm}1 spk & 3 & \xmark & -6.56$\pm$8.08 & 1.38$\pm$0.28 & 0.31$\pm$0.17 & 1.46$\pm$0.54 & 131.9 & 113.1 & {130.7} & {111.9} \\
            \hspace{2mm}2 spk & 3 & \xmark & 1.62$\pm$6.13 & 1.69$\pm$0.37 & 0.50$\pm$0.17 & 2.16$\pm$0.45 & 74.2 & 65.0 & {75.1} & {64.1} \\
            \hspace{2mm}3 spk & 3 & \xmark & 6.09$\pm$6.00 & 1.99$\pm$0.39 & 0.62$\pm$0.15 & \underline{2.43$\pm$0.44} & 46.2 & 44.4 & {48.5} & {44.1} \\
            RAVSS \cite{ravss} & 3 & \xmark & \textbf{9.93$\pm$2.03} & \textbf{2.73$\pm$0.29} & \textbf{0.71$\pm$0.09} & 1.91$\pm$0.38 & \underline{20.6} & \underline{20.0} & {\underline{24.8}} & {\underline{22.4}} \\
            SSNAPS (ours) & 3 & \checkmark & \underline{7.66$\pm$3.40} & \underline{2.28$\pm$0.41} & \underline{0.67$\pm$0.10} & \textbf{2.74$\pm$0.45} & \textbf{19.0} & \textbf{19.6} & {\textbf{23.0}} & {\textbf{21.8}} \\
            \bottomrule
        \end{tabular}%
      \end{sc}
  \end{center}
  \vskip -0.1in
\end{table*}

\subsubsection{Pseudo Ground-Truth Transcriptions}
As the evaluation dataset does not provide human-annotated ground-truth transcriptions, we adopt a pseudo-ground-truth protocol to compute \ac{WER}. Specifically, for each utterance, we obtain a reference transcription by applying the same \ac{ASR} model to the corresponding clean speech signal. This transcription is then treated as the reference when computing \ac{WER} for the enhanced or separated signal produced by the evaluated algorithm. {We also run cross-\ac{ASR} tests to see whether \ac{WER}-based rankings change when one \ac{ASR} model transcribes clean signals and a different model transcribes the estimated speech.}

Both \ac{ASR} models used in this work are state-of-the-art systems that achieve low \ac{WER} on clean speech, making them suitable for generating reliable pseudo ground-truth transcriptions.

\section{Results}
\label{sec:results}
This section presents experimental results comparing SSNAPS with leading baselines for speech separation, speech enhancement, and off-screen speech separation. In addition, the suitability of the recovered noise signal for acoustic scene detection is verified.

\subsection{Speech Separation and Enhancement Results}
The speech separation results for mixtures of 1, 2, and 3 speakers with noise, using the VoxCeleb2+DNS and VoxCeleb2+DCASE combinations, are presented in Tables~\ref{tb:dns1} and~\ref{tb:dcase}. For each case, we evaluate the performance of FlowAVSE when trained with different numbers of speakers, in order to assess whether a single FlowAVSE model can generalise across scenarios, as demonstrated by SSNAPS and RAVSS.

Several trends can be observed in the tables. {Firstly, with respect to \ac{WER}, SSNAPS achieves the best (lowest) score on both \ac{ASR} models in all the 2- and 3-speaker conditions. For 1-speaker mixtures, except for two cases where FlowAVSE achieves the lowest \ac{WER}, SSNAPS outperforms all of the baselines. Note that this result holds regardless of which ASR model transcribed the ground-truth and the recovered speech signals.} This is particularly remarkable, given that SSNAPS never encounters speech-and-noise mixtures during training. Nevertheless, it still surpasses supervised frameworks that were exposed to the tested mixture type during training. In terms of quality and intelligibility metrics, while either of the supervised baselines (RAVSS, FlowAVSE) often achieves better results, SSNAPS performs competitively across all cases even though it was not trained on speech-and-noise mixtures. Sample audio files exemplifying the reconstructed speech quality can be found on the demo page.\textsuperscript{\ref{fn:demo}}

The second trend evident from the tables is that FlowAVSE is highly sensitive to mismatches in the number of sources between the training and test conditions. For instance, in the two-speaker mixtures, we see that when FlowAVSE was trained on three or a single speaker, its performance across all metrics declined. This phenomenon is especially severe when the model is trained on a single speaker but tested on two.

Thirdly, we observe that FlowAVSE's performance drop is more pronounced as the number of speakers increases, even under matching train-test conditions. As mentioned in \cite{ravss}, this can be attributed to the fact that given the visual features $V_i$, FlowAVSE extracts only the $i$-th speaker. Consequently, when $K$ speakers are present, they are all estimated sequentially and thus independently. In contrast, SSNAPS and RAVSS take a more holistic approach, accounting for all sources during the separation process.

Eventually, comparing Tables~\ref{tb:dns1} and~\ref{tb:dcase} indicates that for most cases, all methods exhibit degraded \ac{SI-SDR} when trained on DNS noise but tested on mixtures with DCASE noise. However, they still perform relatively well considering the domain shift with respect to the noise dataset. We postulate that since DNS is a large and highly rich dataset, it overlaps with DCASE which only comprises ten noise classes.

Remarkably, we note that for 3-speaker+noise mixtures, SSNAPS performs slightly better for DCASE noise compared with Table~\ref{tb:dns1}. This may be explained by the fact that DCASE noise signals tend to be predominantly stationary, whereas DNS noise recordings contain both stationary and non-stationary noise sources. Therefore, at an SNR of 15 dB, it might be easier for the speech and noise priors to pick up their respective components, even though the noise prior was trained on the DNS dataset. 

{Table \ref{tb:enhance_bands} provides a breakdown of the performance on 1-speaker mixtures per input \ac{SNR} range for VoxCeleb2+DNS. For FlowAVSE, only the model trained on a single speaker is considered. As expected, the SI-SDR obtained by all methods improve with the input \ac{SNR}, while the improvement over the mixture shrinks. In most cases, FlowAVSE achieves the best performance with respect to SI-SDR, PESQ and ESTOI, and SSNAPS is slightly worse but still close to RAVSS. For NISQA, FlowAVSE wins irrespective of the input \ac{SNR} band and SSNAPS is the runner-up. In terms of \ac{WER}, for input \ac{SNR} values in the range [-5,5]~dB, SSNAPS achieves the best or second-best WER. Across all methods, \ac{WER} improvement saturates with increased input \ac{SNR}.}

\begin{table*}[t]
  \caption{Speech separation performance for VoxCeleb2+DCASE 1, 2 and 3 speakers+noise mixtures. WER-N and WER-W denote the \ac{WER} obtained by the NVIDIA and Whisper \acp{ASR}, respectively.}
  \label{tb:dcase}
  \begin{center}
      \begin{sc}
        \begin{tabular}{lcccccccc} 
        \toprule
        Method & \#Spk & Unsup. & SISDR$\uparrow$ & PESQ$\uparrow$ & ESTOI$\uparrow$ & NISQA$\uparrow$ & WER-N$\downarrow$ & WER-W$\downarrow$ \\
        \midrule
        Mixture & 1 & - & 2.98$\pm$4.34 & 2.04$\pm$0.57 & 0.56$\pm$0.16 & 1.34$\pm$0.65 & 14.1 & 13.6 \\
        \hdashline
        FlowAVSE \cite{flowavse} & 1 & \xmark & \textbf{11.66$\pm$4.02} & \underline{2.79$\pm$0.69} & \textbf{0.76$\pm$0.12} & \textbf{3.91$\pm$0.84} & \textbf{11.5} & \underline{12.0} \\
        RAVSS \cite{ravss} & 1 & \xmark & \underline{11.34$\pm$4.10} & \textbf{2.86$\pm$0.55} & \underline{0.75$\pm$0.13} & 2.45$\pm$0.57 & 13.2 & 12.1 \\
        SSNAPS (ours) & 1 & \checkmark & 9.14$\pm$3.88 & 2.49$\pm$0.56 & 0.69$\pm$0.13 & \underline{2.66$\pm$0.54} & \underline{12.0} & \textbf{11.4} \\
        \midrule
        Mixture & 2 & - & -2.58$\pm$2.44 & 1.51$\pm$0.22 & 0.35$\pm$0.10 & 1.06$\pm$0.47 & 91.9 & 78.9 \\
        \hdashline
        FlowAVSE \cite{flowavse} & 2 & \xmark & \underline{6.80$\pm$4.84} & 2.07$\pm$0.46 & \underline{0.61$\pm$0.14} & \textbf{3.51$\pm$0.70} & 29.5 & 26.9 \\
        RAVSS \cite{ravss} & 2 & \xmark & \textbf{7.92$\pm$2.85} & \textbf{2.33$\pm$0.38} & \textbf{0.62$\pm$0.12} & 1.97$\pm$0.48 & \underline{24.3} & \underline{22.4} \\
        SSNAPS (ours) & 2 & \checkmark & 6.35$\pm$3.58 & \underline{2.09$\pm$0.42} & \underline{0.61$\pm$0.12} & \underline{2.55$\pm$0.50} & \textbf{22.7} & \textbf{21.6} \\
        \midrule
        Mixture & 3 & - & -3.08$\pm$0.20 & 1.43$\pm$0.14 & 0.34$\pm$0.07 & 0.80$\pm$0.35 & 152.9 & 170.2 \\
        \hdashline
        FlowAVSE \cite{flowavse} & 3 & \xmark & 6.08$\pm$6.48 & 2.03$\pm$0.41 & 0.62$\pm$0.16 & \underline{2.44$\pm$0.46} & 37.9 & 35.0 \\
        RAVSS \cite{ravss} & 3 & \xmark & \textbf{8.30$\pm$2.37} & \textbf{2.36$\pm$0.29} & \underline{0.66$\pm$0.09} & 1.92$\pm$0.40 & \underline{19.9} & \underline{19.9} \\
        SSNAPS (ours) & 3 & \checkmark & \underline{7.84$\pm$3.55} & \underline{2.34$\pm$0.42} & \textbf{0.68$\pm$0.11} & \textbf{2.80$\pm$0.46} & \textbf{18.0} & \textbf{17.3} \\
        \bottomrule
        \end{tabular}
      \end{sc}
  \end{center}
  \vskip -0.1in
\end{table*}

\begin{table*}[t]
  \caption{{Performance per input \ac{SNR} range on speech enhancement for VoxCeleb2+DNS.}}
  \label{tb:enhance_bands}
  \begin{center}
      \begin{sc}
        \begin{tabular}{clccccccc} 
        \toprule
        SNR$_\text{in}$ & Method & Unsup. & SISDR$\uparrow$ & PESQ$\uparrow$ & ESTOI$\uparrow$ & NISQA$\uparrow$ & WER-N$\downarrow$ & WER-W$\downarrow$ \\
        \midrule
        \multirow{4}{*}{{[-5,0]}} & Mixture & - & -2.39$\pm$1.40 & 1.64$\pm$0.39 & 0.52$\pm$0.17 & 1.17$\pm$0.69 & 20.6 & 19.6 \\
        \cdashline{2-9}
        & FlowAVSE \cite{flowavse} &  \xmark & \textbf{11.80$\pm$3.96} & \textbf{2.66$\pm$0.69} & \textbf{0.77$\pm$0.13} & \textbf{3.84$\pm$0.76} & \underline{17.6} & \textbf{14.6} \\
        & RAVSS \cite{ravss} & \xmark & \underline{9.78$\pm$3.42} & \underline{2.59$\pm$0.51} & \underline{0.71$\pm$0.13} & 2.18$\pm$0.56 & 18.4 & 17.4 \\
        & SSNAPS (ours) & \checkmark & 8.97$\pm$4.05 & 2.46$\pm$0.58 & \underline{0.71$\pm$0.14} & \underline{2.44$\pm$0.62} & \textbf{13.1} & \underline{15.2} \\
        \midrule
        \multirow{4}{*}{{[0,5]}} & Mixture & - & 2.65$\pm$1.42 & 1.89$\pm$0.39 & 0.60$\pm$0.14 & 1.26$\pm$0.65 & 13.8 & 12.7 \\
        \cdashline{2-9}
        & FlowAVSE \cite{flowavse} & \xmark & \textbf{13.85$\pm$4.27} & \underline{2.93$\pm$0.74} & \textbf{0.81$\pm$0.11} & \textbf{3.89$\pm$0.85} & \textbf{12.1} & \underline{12.3} \\
        & RAVSS \cite{ravss} & \xmark & \underline{12.34$\pm$3.66} & \textbf{2.94$\pm$0.48} & \underline{0.76$\pm$0.11} & 2.43$\pm$0.50 & 13.9 & 14.6 \\
        & SSNAPS (ours) & \checkmark & 11.32$\pm$3.77 & 2.64$\pm$0.58 & 0.73$\pm$0.13 & \underline{2.55$\pm$0.64} & \underline{12.6} & \textbf{11.9} \\
        \midrule
        \multirow{4}{*}{{[5,10]}} & Mixture & - & 7.58$\pm$1.54 & 2.34$\pm$0.60 & 0.73$\pm$0.13 & 1.67$\pm$0.72 & 6.8 & 8.3 \\
        \cdashline{2-9}
        & FlowAVSE \cite{flowavse} & \xmark & \textbf{16.64$\pm$4.06} & \textbf{3.43$\pm$0.59} & \textbf{0.88$\pm$0.08} & \textbf{3.78$\pm$0.89} & 8.0 & \textbf{7.4} \\
        & RAVSS \cite{ravss} & \xmark & \underline{14.20$\pm$4.47} & \underline{3.25$\pm$0.51} & \underline{0.83$\pm$0.09} & 2.75$\pm$0.55 & \textbf{7.4} & \underline{7.7} \\
        & SSNAPS (ours) & \checkmark & 13.52$\pm$4.16 & 3.00$\pm$0.62 & 0.80$\pm$0.10 & \underline{2.82$\pm$0.64} & \underline{7.9} & 8.4 \\
        \bottomrule
        \end{tabular}
      \end{sc}
  \end{center}
  \vskip -0.1in
\end{table*}

\begin{table*}[t]
  \caption{Separation results for mixtures of 2 speakers (one on-screen and one off-screen)+noise from VoxCeleb2. Visual features are available for the on-screen speaker but not for the off-screen counterpart.}
  \label{tb:offscreen2}
  \begin{center}
      \begin{sc}
        \begin{tabular}{lcccccccc} 
        \toprule
        Method & on-screen & Unsup. & SISDR$\uparrow$ & PESQ$\uparrow$ & ESTOI$\uparrow$ & NISQA$\uparrow$ & WER-N$\downarrow$ & WER-W$\downarrow$ \\
        \midrule
        Mixture & \checkmark & - & -2.54$\pm$2.64 & 1.48$\pm$0.21 & 0.38$\pm$0.11 & 1.13$\pm$0.54 & 93.0 & 84.7 \\
        \hdashline
        RAVSS \cite{ravss} & \checkmark & \xmark & \textbf{8.46$\pm$3.79} & \textbf{2.42$\pm$0.42} & \textbf{0.65$\pm$0.13} & 2.02$\pm$0.46 & \textbf{25.1} & \textbf{24.4}\\
        SSNAPS (ours) & \checkmark & \checkmark & \underline{6.24$\pm$4.16} & \underline{2.17$\pm$0.50} & \underline{0.64$\pm$0.13} & \textbf{2.64$\pm$0.59} & \underline{25.5} & \underline{26.1} \\
        \hspace{2mm}-w/o $\gL_{\textrm{ct-ss}}$ & \checkmark & \checkmark & 6.07$\pm$4.53 & 2.11$\pm$0.50 & 0.62$\pm$0.15 & \underline{2.46$\pm$0.49} & 27.3 & 26.9 \\
        \midrule
        Mixture & \xmark & - & -2.55$\pm$2.53 & 1.47$\pm$0.20 & 0.38$\pm$0.11 & 1.13$\pm$0.54 & 95.4 & 94.2 \\
        \hdashline
        RAVSS \cite{ravss} & \xmark & \xmark & \textbf{8.43$\pm$3.70} & \textbf{2.37$\pm$0.41} & \textbf{0.64$\pm$0.13} & 1.97$\pm$0.46 & \textbf{29.0} & \textbf{32.1} \\
        SSNAPS (ours) & \xmark & \checkmark & \underline{4.56$\pm$4.77} & \underline{1.94$\pm$0.46} & \underline{0.59$\pm$0.15} & \textbf{2.41$\pm$0.66} & \underline{38.4} & \underline{39.8} \\
        \hspace{2mm}-w/o $\gL_{\textrm{ct-ss}}$ & \xmark & \checkmark & 3.48$\pm$5.32 & 1.76$\pm$0.46 & 0.53$\pm$0.17 & 2.29$\pm$0.57 & 55.7 & 54.6 \\
        \bottomrule
        \end{tabular}
      \end{sc}
  \end{center}
  \vskip -0.1in
\end{table*}

\subsection{Off-Screen Speech Separation Results}
The speech separation results for VoxCeleb2+DNS 2-speaker+noise mixtures, when visual features are available only for the on-screen speakers, are provided in Table \ref{tb:offscreen2}. The results are reported for RAVSS and SSNAPS. Recall that as an essential part of its training process, RAVSS was randomly tasked with mixtures in which only some of the speakers had visual cues. As an ablation study, we assess the performance of SSNAPS with $\gL_{\textrm{os}} = \gL_{\text{rec}} 
+ g_{\textrm{ct-ss}}\gL_{\textrm{ct-ss}}$ and also with just $\gL_{\textrm{os}} = \gL_{\text{rec}}$ to assert the importance of our proposed loss function $\gL_{\textrm{ct-ss}}$. The results for the on-screen and off-screen speakers are shown separately to provide clearer insight into how each method handles each source.

\begin{table*}[t]
  \caption{Separation results for mixtures of 3 speakers+noise from VoxCeleb2. Visual features are available for the two on-screen speakers but not for the off-screen counterpart. The metrics for the on-screen speakers are averaged over the corresponding speakers.}
  \label{tb:offscreen3}
  \begin{center}
      \begin{sc}
        \begin{tabular}{lcccccccc} 
        \toprule
        Method & on-screen & Unsup. & SISDR$\uparrow$ & PESQ$\uparrow$ & ESTOI$\uparrow$ & NISQA$\uparrow$ & WER-N$\downarrow$ & WER-W$\downarrow$ \\
        \midrule
        Mixture & \checkmark & - & -3.08$\pm$0.10 & 1.44$\pm$0.11 & 0.33$\pm$0.04 & 0.87$\pm$0.39 & 172.3 & 171.0 \\
        RAVSS \cite{ravss} & \checkmark & \xmark & \textbf{7.71$\pm$3.05} & \textbf{2.32$\pm$0.26} & \underline{0.64$\pm$0.08} & 1.92$\pm$0.38 & \underline{22.5} & \underline{24.9}\\
        SSNAPS (ours) & \checkmark & \checkmark & \underline{6.17$\pm$3.78} & \underline{2.16$\pm$0.42} & \textbf{0.65$\pm$0.11} & \textbf{2.78$\pm$0.43} & \textbf{21.7} & \textbf{24.4} \\
        \hspace{2mm}-w/o $\gL_{\textrm{ct-ss}}$ & \checkmark & \checkmark & 5.96$\pm$3.69 & 2.10$\pm$0.38 & \textbf{0.65$\pm$0.10} & \underline{2.71$\pm$0.44} & 29.4 & 27.6 \\
        \hdashline
        Mixture & \xmark & - & -3.08$\pm$0.14 & 1.44$\pm$0.14 & 0.33$\pm$0.07 & 0.87$\pm$0.39 & 208.0 & 205.8 \\
        RAVSS \cite{ravss} & \xmark & \xmark & \textbf{7.12$\pm$5.27} & \textbf{2.23$\pm$0.34} & \textbf{0.62$\pm$0.13} & 1.85$\pm$0.37 & \textbf{28.5} & \textbf{27.0} \\
        SSNAPS (ours) & \xmark & \checkmark & \underline{4.28$\pm$4.85} & \underline{1.81$\pm$0.42} & \underline{0.56$\pm$0.15} & \textbf{2.71$\pm$0.40} & \underline{55.1} & \underline{55.1} \\
        \hspace{2mm}-w/o $\gL_{\textrm{ct-ss}}$ & \xmark & \checkmark & 3.71$\pm$5.32 & 1.71$\pm$0.37 & 0.55$\pm$0.15 & \underline{2.59$\pm$0.41} & 60.5 & 58.7 \\
        \bottomrule
        \end{tabular}
      \end{sc}
  \end{center}
  \vskip -0.1in
\end{table*}

As Table~\ref{tb:offscreen2} shows, RAVSS has the best scores in all metrics except for NISQA. In our informal listening tests, we noticed that RAVSS output indeed lacked naturalness.
Across all methods, it is evident that the on-screen speaker can be more effectively isolated, even for the ablated SSNAPS. It can also be seen that our proposed loss, $\gL_{\textrm{ct-ss}}$, significantly improved SSNAPS results for the off-screen speaker. Moreover, it also slightly improved all metrics for the on-screen speaker. Sample speech recordings are available on the demo page.\textsuperscript{\ref{fn:demo}}

Separation performance for off-screen speech separation for 3-speaker+noise mixtures and an ablation study for SSNAPS is reported in Table~\ref{tb:offscreen3}. Recall that two speakers have visual cues, and the third one does not. The metrics presented in Table \ref{tb:offscreen3} for the on-screen speakers are averaged over the two speakers. For the on-screen speaker, we can see that SSNAPS achieves the best ESTOI, NISQA, and WER. Without using $\gL_{\textrm{ct-ss}}$, the \ac{WER} significantly drops. For the off-screen speaker, RAVSS outperforms by a large margin, albeit lacking in naturalness, as reflected by its low NISQA score. For SSNAPS, incorporating $\gL_{\textrm{ct-ss}}$ improves its performance, but to a lesser extent than for on-screen speakers.  

{The gap between SSNAPS and RAVSS on the off-screen speaker (e.g., 4.56 vs.\ 8.43\,dB SI-SDR in Table~\ref{tb:offscreen2}) is considerably larger than the on-screen speakers, for which SSNAPS remains relatively close to or even surpasses RAVSS in several metrics (Table~\ref{tb:offscreen3}). It suggests that the missing visual conditioning is the source of the performance drop. We attribute this to two factors. First, RAVSS, unlike SSNAPS, is trained in a supervised manner, end-to-end, on mixtures with randomly missing visual cues, and can therefore learn to separate a video-less speaker from paired data. Second, without visual guidance, the speech prior utilised by SSNAPS is considerably less constraining, and becomes susceptible to attributing speech of speaker to a different one. }

{Notably, by comparing Table \ref{tb:dns1} with Tables \ref{tb:offscreen2}, \ref{tb:offscreen3}, we observe that for SSNAPS the performance of both on- and off-screen speakers worsens with the respect to the setup with visual features present for all speakers. It is hence deduced that the lack of video for one speaker also affects the ability to correctly separate the others when their characteristics in the frequency domain resemble each other. The proposed loss $\gL_{\textrm{ct-ss}}$ discourages leakage but does not encode speaker identity. We thus conclude that further research is required for this task.}

\begin{table*}[t]
  \caption{{Audio-only speech enhancement on VoxCeleb2+DNS. SSNAPS-A denote SSNAPS without visual features.}}
  \label{tb:audio_only}
  \begin{center}
      \begin{sc}
        \begin{tabular}{clccccccc} 
            \toprule
            SNR$_\text{in}$ & Method & Unsup. & SISDR$\uparrow$ & PESQ$\uparrow$ & ESTOI$\uparrow$ & NISQA$\uparrow$ & WER-N$\downarrow$ & WER-W$\downarrow$ \\
            \midrule
            \multirow{5}{*}{{[-5,0]}} & Mixture & - & -2.39$\pm$1.41 & 1.64$\pm$0.38 & 0.51$\pm$0.17 & 1.17$\pm$0.69 & 20.6 & 19.6 \\
            & SSNAPS (ours) & \checkmark & 8.97$\pm$4.05 & 2.46$\pm$0.58 & 0.71$\pm$0.14 & 2.44$\pm$0.62 & 13.1 & 15.2 \\
            \cdashline{2-9}
            & SGMSE \cite{sgmse} & \xmark & \underline{3.84$\pm$5.17} & \underline{2.18$\pm$0.58} & \underline{0.63$\pm$0.17} & \textbf{2.66$\pm$0.92} & 30.4 & 30.3 \\
            & UDiffSE \cite{nmf1} & \checkmark & -1.03$\pm$3.12 & 1.70$\pm$0.42 & 0.50$\pm$0.17 & 1.66$\pm$0.85 & \textbf{26.8} & \textbf{24.1} \\
            & SSNAPS-A (ours) & \checkmark & \textbf{7.68$\pm$4.48} & \textbf{2.26$\pm$0.62} & \textbf{0.66$\pm$0.17} & \underline{2.54$\pm$0.61} & \underline{28.5} & \underline{28.1} \\
            \midrule
            \multirow{5}{*}{{[0,5]}} & Mixture & - & 2.65$\pm$1.42 & 1.89$\pm$0.39 & 0.60$\pm$0.15 & 1.26$\pm$0.65 & 13.8 & 12.7 \\
            & SSNAPS (ours) & \checkmark & 11.32$\pm$3.77 & 2.64$\pm$0.58 & 0.73$\pm$0.13 & 2.55$\pm$0.64 & 12.6 & 11.9 \\
            \cdashline{2-9}
            & SGMSE \cite{sgmse} & \xmark & \underline{8.90$\pm$4.03} & \textbf{2.60$\pm$0.55} & \textbf{0.71$\pm$0.14} & \textbf{3.08$\pm$0.76} & 20.7 & 21.3 \\
            & UDiffSE \cite{nmf1} & \checkmark & 4.94$\pm$2.51 & 2.00$\pm$0.40 & 0.60$\pm$0.13 & 1.93$\pm$0.75 & \textbf{15.8} & \textbf{16.8} \\
            & SSNAPS-A (ours) & \checkmark & \textbf{10.11$\pm$4.18} & \underline{2.49$\pm$0.60} & \underline{0.70$\pm$0.14} & \underline{2.62$\pm$0.56} & \underline{18.7} & \underline{19.0} \\
            \midrule
            \multirow{5}{*}{{[5,10]}} & Mixture & - & 7.58$\pm$1.54 & 2.35$\pm$0.61 & 0.73$\pm$0.13 & 1.67$\pm$0.72 & 6.8 & 8.3 \\
            & SSNAPS (ours) & \checkmark & 13.52$\pm$4.16 & 3.00$\pm$0.62 & 0.80$\pm$0.10 & 2.82$\pm$0.64 & 7.9 & 8.4 \\
            \cdashline{2-9}
            & SGMSE \cite{sgmse} & \xmark & \underline{12.14$\pm$3.98} & \textbf{2.98$\pm$0.58} & \textbf{0.79$\pm$0.10} & \textbf{3.36$\pm$0.71} & \underline{11.3} & \underline{12.3} \\
            & UDiffSE \cite{nmf1} & \checkmark & 8.90$\pm$2.39 & 2.44$\pm$0.54 & 0.72$\pm$0.11 & 2.47$\pm$0.74 & \textbf{8.7} & \textbf{9.1} \\
            & SSNAPS-A (ours) & \checkmark & \textbf{12.33$\pm$4.68} & \underline{2.83$\pm$0.65} & \underline{0.77$\pm$0.12} & \underline{2.85$\pm$0.59} & 12.6 & \underline{12.3} \\
            \midrule
            \multirow{5}{*}{{[-5,10]}} & Mixture & - & 2.61$\pm$4.28 & 1.95$\pm$0.55 & 0.61$\pm$0.17 & 1.36$\pm$0.72 & 13.8 & 13.5 \\
            & SSNAPS (ours) & \checkmark & 11.83$\pm$4.39 & 2.73$\pm$0.63 & 0.76$\pm$0.13 & 2.61$\pm$0.65 & 11.3 & 11.8 \\
            \cdashline{2-9}
            & SGMSE \cite{sgmse} & \xmark & \underline{8.32$\pm$5.56} & \textbf{2.59$\pm$0.65} & \textbf{0.71$\pm$0.15} & \textbf{3.04$\pm$0.85} & 20.9 & 21.3 \\
            & UDiffSE \cite{nmf1} & \checkmark & 4.29$\pm$4.85 & 2.04$\pm$0.55 & 0.61$\pm$0.17 & 2.02$\pm$0.85 & \textbf{17.2} & \textbf{16.7} \\
            & SSNAPS-A (ours) & \checkmark & \textbf{10.04$\pm$4.82} & \underline{2.52$\pm$0.67} & \textbf{0.71$\pm$0.15} & \underline{2.67$\pm$0.60} & \underline{19.9} & \underline{19.8} \\
            \bottomrule
        \end{tabular}%
      \end{sc}
  \end{center}
  \vskip -0.1in
\end{table*}

\subsection{{Audio-Only Speech Enhancement}}
{Table~\ref{tb:audio_only} reports single-speaker enhancement without visual cues. A comparison of SSNAPS applied with and without visual cues shows the contribution of the visual input. The visual information improves SI-SDR by up to 1.2-1.8~dB, PESQ by 0.15-0.21 and ESTOI by 0.03-0.05. Its main effect is on semantic content, as it lowers \ac{WER} by roughly a third to a half, most strongly at low SNR (e.g., 28.5 to 13.1 \ac{WER}-N at [-5,0]~dB). NISQA is marginally lower with video (by at most 0.10). This highlights value of the visual input.}

{Without visual cues, SSNAPS remains competitive with audio-only methods. Compared with the unsupervised UDiffSE, it is considerably better in SI-SDR, PESQ, ESTOI and NISQA in all ranges, although UDiffSE attains a lower \ac{WER}. Compared to SGMSE, SSNAPS has higher SI-SDR in all ranges and a lower \ac{WER} in all but the [5,10]~dB range, whereas SGMSE has higher NISQA and, except at [-5,0]~dB, higher PESQ.}

{Interestingly, all the audio-based methods have higher \ac{WER} than the unprocessed mixture. One possible explanation is that their generative nature introduces unintelligible hallucinated speech. UDiffSE attains the lowest \ac{WER} among the audio-only methods. however, it does so with only modest enhancement, i.e., it modifies the mixture less, whereas methods with larger perceptual gains also incur a larger \ac{WER} degradation. Modern \acp{ASR} like Whisper are robust to additive noise and sensitive to artifacts, which is why a method that changes little can win on \ac{WER}.}

\begin{table*}
  \caption{Sampling efficiency for SSNAPS and the competing methods is presented for the separation of 1, 2, and 3 speech signals and noise. The number of Langevin iterations, the number of ODE steps $N_{\text{ODE}}$, and Annealing steps $N_A$ are pertinent only to SSNAPS.}
  \label{tb:runtime}
  \begin{center}
    \begin{small}
      \begin{sc}
        \begin{tabular}{lcccccc} 
        \toprule
        Method & \#Spk & Langevin Steps & ODE Steps & Annealing Steps & NFE & Seconds \\
        \midrule
        FlowAVSE & 1 & - & - & - & 1 & 0.06 \\
        FlowAVSE & 2 & - & - & - & 1 & 0.12 \\
        FlowAVSE & 3 & - & - & - & 1 & 0.18 \\
        \midrule
        RAVSS  & 1 & - & - & - & 1 & 0.07 \\
        RAVSS  & 2 & - & - & - & 1 & 0.12 \\
        RAVSS  & 3 & - & - & - & 1 & 0.18 \\
        \midrule
        DAVSS-NM & 2 & - & - & - & 1600 & 110 \\
        \midrule
        SSNAPS & 1 & 50 & 2 & 300 & 1200 & 72 \\
        SSNAPS & 2 & 100 & 2 & 300 & 1200 & 110 \\
        SSNAPS & 3 & 100 & 2 & 400 & 1600 & 160 \\
        \bottomrule
        \end{tabular}
      \end{sc}
    \end{small}
  \end{center}
  \vskip -0.1in
\end{table*}

\begin{table*}[t]
  \caption{{Ablation on $N_\text{A}$ and $N_\text{MC}$  as evaluated on 2-speaker mixtures from VoxCeleb2+DNS. The asterisk symbol denotes the setting used by SSNAPS as reported in Table~ \ref{tb:imp}.}}
  \label{tb:ablation}
  \begin{center}
      \begin{sc}
        \begin{tabular}{lcccccccc} 
            \toprule
             & SISDR$\uparrow$ & PESQ$\uparrow$ & ESTOI$\uparrow$ & NISQA$\uparrow$ & WER-N$\downarrow$ & WER-W$\downarrow$ & NFE & Seconds \\
            \midrule
            Mixture & -2.74$\pm$3.01 & 1.47$\pm$0.23 & 0.38$\pm$0.11 & 1.16$\pm$0.57 & 90.4 & 93.6 & - & - \\
            \midrule
            $N_\text{MC}=10$ & \underline{7.60$\pm$3.76} & 2.22$\pm$0.50 & \textbf{0.66$\pm$0.12} & \textbf{2.71$\pm$0.62} & 19.5 & \textbf{17.8} & 1200 & 55 \\
            $N_\text{MC}=50$ & 7.57$\pm$3.95 & 2.24$\pm$0.50 & \textbf{0.66$\pm$0.13} & \underline{2.54$\pm$0.67} & \underline{19.1} & 21.2 & 1200 & 80 \\
            $N_\text{MC}=100$* & \textbf{7.68$\pm$3.75} & \textbf{2.26$\pm$0.51} & \textbf{0.66$\pm$0.12} & 2.48$\pm$0.68 & \textbf{18.2} & \underline{18.1} & 1200 & 110 \\
            $N_\text{MC}=200$ & 7.42$\pm$4.07 & \underline{2.25$\pm$0.49} & \textbf{0.66$\pm$0.13} & 2.40$\pm$ 0.70 & 20.1 & 21.0 & 1200 & 172 \\
            \midrule
            $N_\text{A}=100$ & 6.68$\pm$3.82 & 2.19$\pm$0.46 & \underline{0.64$\pm$0.13} & 2.17$\pm$0.76 & 19.7 & 19.7 & 400 & 37 \\
            $N_\text{A}=200$ & 7.20$\pm$3.87 & \underline{2.24}$\pm$0.48 & \textbf{0.66$\pm$0.12} & 2.41$\pm$0.72 & \underline{18.3} & \underline{17.6} & 800 & 74 \\
            $N_\text{A}=300$* & \underline{7.68$\pm$3.75} & \textbf{2.26$\pm$0.51} & \textbf{0.66$\pm$0.12} & \underline{2.48$\pm$0.68} & \textbf{18.2} & 18.1 & 1200 & 110 \\
            $N_\text{A}=400$ & \textbf{7.83$\pm$3.89} & \textbf{2.26$\pm$0.52} & \textbf{0.66$\pm$0.13} & \textbf{2.51$\pm$0.65} & \textbf{18.2} & \textbf{17.3} & 1600 & 175 \\
            \bottomrule
        \end{tabular}%
      \end{sc}
  \end{center}
  \vskip -0.1in
\end{table*}

\subsection{Acoustic Scene Detection}
\begin{table}[t]
  \caption{Acoustic scene detection results for 1 speaker+noise mixtures from VoxCeleb2+DCASE.}
  \label{tb:scene_detect}
  \begin{center}
      \begin{sc}
        \begin{tabular}{ccccccc} 
        \toprule
         & \multicolumn{3}{c}{Mixture} & \multicolumn{3}{c}{SSNAPS} \\
        \cmidrule(lr){2-4} \cmidrule(lr){5-7}
        SNR & 3 & 10 & Fused & 3 & 10 & Fused \\
        \midrule
        -10 & 77.6 & 65.8 & 64.6 & 82.8 & 74.7 & 74.8 \\
        -5 & 70.2 & 50.4 & 50.2 & 82.0 & 68.2 & 69.0 \\
        0 & 65.8 & 35.2 & 37.8 & 78.6 & 59.6 & 61.0 \\
        5 & 62.0 & 25.8 & 32.1 & 70.2 & 45.6 & 44.2 \\
        \bottomrule
        \end{tabular}
      \end{sc}
  \end{center}
  \vskip -0.1in
\end{table}
Beyond extracting the individual speech sources, SSNAPS also provides an estimate of the ambient noise signal. To examine how well this estimate can be used for acoustic scene detection, we compare the detection accuracy achieved with the reconstructed noise against that obtained using the mixture signal on the DCASE 2020 Challenge \cite{dcase}. In this challenge, 10-second environmental noise recordings from the DCASE dataset must be assigned to one of 10 classes, with each class corresponding to a distinct environment. 

We assess the classification accuracy by employing \cite{dcase_classifier}, a classifier designed to address the DCASE Challenge. This method proposes to further group the ten noise classes into three hyperclasses. Then, two classifiers are trained: one for the original classes and another auxiliary classifier for the hyperclasses, which assists the 10-class classifier. Finally, the classification results are fused and translated to a final assignment of the noise recording to one of the original ten classes.

We evaluate the noise classification results for mixtures of a single speaker and noise at different \ac{SNR} levels.
As \cite{dcase_classifier} assumes pure environmental noise signals, it is expected that overlapping speech will degrade the performance of the classifier.  Consequently, unlike speech enhancement, where higher \ac{SNR} leads to better performance, in acoustic scene detection in the presence of speech, lower \ac{SNR} is preferred. 

The classification results are presented in Table~\ref{tb:scene_detect} for 100 mixtures.  The classification accuracy of the 10-class classifier, the 3-class classifier, and their fusion is computed for the raw mixture and the noise signal recovered by SSNAPS. It is evident that, for both classifiers and their fusion across all \ac{SNR} levels, the noise signal estimated by SSNAPS improves accuracy with respect to the unprocessed signal.

It should be noted that \cite{dcase_classifier} trained the classifiers on 10s-long noise recordings sampled at 44.1 kHz. However, we use only 4s-long mixtures sampled at 16~kHz.
Therefore, for computing the results in Table \ref{tb:scene_detect}, only DCASE noise signals that are accurately classified after truncation to 4 s, downsampling to 16 kHz, and upsampling to 44.1 kHz are considered.

\subsection{Sampling Efficiency}
\label{app:runtime}
Table~\ref{tb:runtime} reports the runtime and the \ac{NFE} of SSNAPS and the baselines for all mixture types, for separating 4s-long mixtures. All experiments are conducted on a single NVIDIA H200 GPU. Unsurprisingly, the supervised methods FlowAVSE and RAVSS achieve the fastest inference and the lowest \ac{NFE}. The long runtime of DAVSS-NM and SSNAPS exemplifies one of the main weaknesses of inverse sampling methods: their slow runtime. We note that, since SSNAPS uses classifier-free guidance, its \ac{NFE} value is, in practice, $2N_{\text{ODE}}N_A$. It is also apparent that using more Langevin updates increases SSNAPS' runtime due to the additional gradient computations.

\subsection{{Ablation Study}}
{To examine the trade-off between performance and computational cost, Table~\ref{tb:ablation} reports an ablation on $N_\text{A}$ and $N_\text{MC}$ for 100 VoxCeleb2+DNS 2-speaker mixtures. Naturally, the runtime decreases with $N_\text{A}$ and $N_\text{MC}$. Performance steadily improves with $N_\text{A}$. Even with $N_\text{A}=100$, three times fewer \acp{NFE} than the default, SSNAPS achieves significant improvement over the mixture signal.}

{The performance is largely insensitive to $N_\text{MC}$. SI-SDR varies within 0.3~dB for $N_\text{MC}\in[10,200]$, well within the standard deviation, and the \ac{WER} variations are not monotonic, suggesting that they are within evaluation noise. Although we chose $N_\text{MC}=100$ for this paper because it yields slightly better \ac{WER} as measured by WER-N, we can reduce $N_\text{MC}$ substantially, lowering the number of gradient computations, without a notable loss. This finding is also supported by what reported in DAPS++ \cite{chen2026daps++}.}

{While still a few orders of magnitude more computationally demanding than RAVSS and FlowAVSE, it is possible to drastically cut down the runtime by decreasing $N_\text{A}$ and $N_\text{A}$ almost without compromising on performance.}

\section{Conclusion}
We presented a novel unsupervised generative audio-visual method for speech separation and enhancement, which estimates several speech signals and a background noise signal from their mixture. Our method, SSNAPS, utilises both clean speech and noise priors represented by diffusion models. It leverages the inverse-problem framework to estimate the constituents of the mixture by sampling from the posterior. 
As SSNAPS is an unsupervised technique, it can be applied to various speech processing problems with minimal changes.

We showed that by using expressive data priors combined with a strong inverse sampling method, for mixtures of 1, 2 and 3 speakers and noise mixtures, in most cases SSNAPS achieves the best \ac{WER} score compared to supervised top-tier baselines, while presenting good audio quality.
Moreover, we extended SSNAPS to the highly realistic scenario of off-screen speech separation. Finally, it was shown the ambient noise signal recovered by SSNAPS consistently improves scene acoustic detection with respect to the mixture signal.

{A limitation of SSNAPS, common to diffusion-based inverse solvers, is its computational cost. 
It is orders of magnitude slower than the supervised baselines, which currently precludes real-time use. Our ablation study shows that the number of annealing and Langevin steps can be reduced at a moderate performance cost, but a substantial speed-up will likely require different tools. Future work will therefore explore distilling the speech and noise priors into few-step generators, e.g., consistency models or flow-matching models, as well as more advanced \ac{ODE}/\ac{SDE} solvers.}


\appendix
\section*{Derivation of SSNAPS}
\label{app:ssnaps}
The first stage in a SSNAPS iteration is sampling from the posterior $p(\rvh_0|\rvh_\tau, \rvy)$. Here we devise $p(\rvh_0|\rvh_\tau, \rvy)$ to prove (\ref{eq:posterior}). Using the definition of $\rvh$, we get $p(\rvh_0|\rvh_\tau, \rvy)=p(\rvc_0,\rvn_0|\rvc_\tau, \rvn_\tau, \rvy)$. By applying Bayes' Theorem:
\begin{align}
    p(\rvc_0, &\rvn_0 \mid \rvc_\tau, \rvn_\tau, \rvy) \nonumber\\
    &= \frac{p(\rvy \mid \rvc_0, \rvn_0, \rvc_\tau, \rvn_\tau) p(\rvc_0, \rvn_0,\rvc_\tau, \rvn_\tau)}{p(\rvc_\tau, \rvn_\tau, \rvy)} \nonumber \\
    &=
    \frac{p(\rvy \mid \rvc_0, \rvn_0) p(\rvc_0, \rvn_0\mid\rvc_\tau, \rvn_\tau)}{p(\rvy\mid\rvc_\tau, \rvn_\tau)} \nonumber\\
    &\propto 
    p(\rvy \mid \rvc_0, \rvn_0) p(\rvc_0 \mid \rvc_\tau, \rvn_\tau, \rvn_0)p(\rvn_0 \mid \rvc_\tau, \rvn_\tau).
\end{align}
Now, since all sources are statistically independent, we have
\begin{align}
    p(\rvc_0, \rvn_0 \mid \rvc_\tau, \rvn_\tau, \rvy) 
    &\propto 
    p(\rvy \mid \rvc_0, \rvn_0) p(\rvc_0 \mid \rvc_\tau)p(\rvn_0 \mid \rvn_\tau) \nonumber\\
    &=
    p(\rvy \mid \rvc_0, \rvn_0) p(\rvn_0 \mid \rvn_\tau) \prod_{i=1}^K p(\rvx_0^i \mid \rvx_\tau^i),
\end{align}
which completes the proof.

It is now required to prove that the Langevin update rule
\begin{equation}
    \rvh_0^{(j+1)} = \rvh_0^{(j)} + \eta \nabla_{\rvh_0^{(j)}}\log p(\rvh_0^{(j)}\mid\rvh_{\tau}, \rvy) 
    + \sqrt{2 \eta}\,\boldsymbol{\epsilon}_j
    \label{eq:langevin_h}
\end{equation}
can be decomposed into the $K+1$ separate update rules described by (\ref{eq:langevin_sp}), (\ref{eq:langevin_np}). To achieve this, we plug (\ref{eq:posterior}) into (\ref{eq:langevin_h}):
\begin{align}
    \rvh_0^{(j+1)} &= \rvh_0^{(j)} 
    + \eta \nabla_{\rvh_0^{(j)}}\log p(\rvy \mid \rvc^{(j)}_0, \rvn_0^{(j)}) \nonumber\\
    &+ \eta \sum_{i=1}^K\nabla_{\rvh_0^{(j)}}\log p(\rvx_0^{i^{(j)}} \mid \rvx_\tau^i) \nonumber\\
    &+ \eta \nabla_{\rvh_0^{(j)}}\log p(\rvn_0^{(j)} \mid \rvn_\tau) 
    + \sqrt{2 \eta}\,\boldsymbol{\epsilon}_j.
\end{align}
Recall that $\rvh$ is the concatenation of $\rvc$ and $\rvn$. Therefore, each $d$ consecutive components in $\rvh$ are associated with a single source, such that $\rvh[(i-1)d:id]=\rvx^i$ for $i=1,\ldots,K$ and $\rvh[Kd:(K+1)d]=\rvn$. The gradient operator $\nabla_{\rvh_0^{(j)}}$ returns the derivative with respect to each element of $\rvh$. Consequently, by noting that for all $i=1,\ldots,K$:
\begin{align}
    \nabla_{\rvn}\log p(\rvy \mid \rvc_0, \rvn_0) &= \nabla_{\rvx^i}\log p(\rvy \mid \rvc_0, \rvn_0) \\
    \nabla_{\rvx^i}\log p(\rvn \mid \rvn_\tau) &= \mathbf{0}
    \\
    \nabla_{\rvn}\log p(\rvx_0^i \mid \rvx_\tau^i) &= \mathbf{0}
    \\
    \nabla_{\rvx^j}\log p(\rvx_0^i \mid \rvx_\tau^i) &= \mathbf{0},\; i\neq j.
\end{align}
Hence this proves that (\ref{eq:langevin_h}) is equivalent to the $K+1$ update rules (\ref{eq:langevin_sp}), (\ref{eq:langevin_np}).

\balance
\bibliography{refs}
\bibliographystyle{IEEEtran}

\vfill

\end{document}